\documentclass[12pt]{article}
\usepackage{amsmath}
\usepackage{graphicx}
\usepackage{natbib}
\usepackage{url} 
\usepackage{hyperref}
\newcommand{\blind}{1}
\usepackage{amsmath}
\usepackage{graphicx}
\usepackage{enumerate}
\usepackage{natbib}
\usepackage{url} 

\usepackage{amsmath,amsfonts,amssymb,amsthm}
\usepackage{graphicx,psfrag,epsf}
\usepackage{enumerate}
\usepackage{algorithm,algorithmicx,algpseudocode}
\usepackage{bm}
\usepackage{booktabs}
\usepackage{multirow} 
\usepackage{float}
\usepackage{amsfonts}
\usepackage{algcompatible,amsmath}
\usepackage{caption}
\usepackage[labelformat=simple]{subcaption}
\usepackage{afterpage}
\usepackage{authblk}
\usepackage{dashrule}

\usepackage{amssymb}

\DeclareMathOperator{\Cov}{Cov}
\DeclareMathOperator{\Ex}{E}

\DeclareMathOperator*{\argmin}{argmin}
  
\DeclareMathOperator{\diag}{diag}

\newtheorem{theorem}{Theorem}
\algnewcommand\TO{\item[\textbf{to}]}

\theoremstyle{plain}

\begin{document}

\def\spacingset#1{\renewcommand{\baselinestretch}%
{#1}\small\normalsize} \spacingset{1}


\if1\blind
{
  \title{\bf Sparse and Smooth Functional Data Clustering}
  \author[1]{Fabio Centofanti}
    \author[1]{Antonio Lepore \thanks{Corresponding author. e-mail: \texttt{antonio.lepore@unina.it}}}
     \author[1]{Biagio Palumbo}
  
  \affil[1]{Department of Industrial Engineering, University of Naples Federico II, Piazzale Tecchio 80, 80125, Naples, Italy}
  \date{}
  \maketitle
} \fi

\if0\blind
{
  \bigskip
  \bigskip
  \bigskip
  \begin{center}
    {\LARGE\bf Sparse and Smooth Functional Data Clustering}
\end{center}
  \medskip
} \fi

\bigskip
\begin{abstract}
A   new model-based procedure is developed for  sparse clustering of functional data that aims to classify a sample of curves into homogeneous groups  while jointly detecting the most informative portions of domain.
The proposed method is referred to as sparse and smooth functional clustering (SaS-Funclust) and  relies on a general functional Gaussian mixture
model  whose parameters are estimated by maximizing a  log-likelihood function penalized with a functional adaptive pairwise penalty and a roughness penalty. The former allows identifying the noninformative portion of domain by shrinking the means of separated clusters to some common values, whereas the latter improves the interpretability by imposing some degree of smoothing to the estimated cluster means. The  model is estimated  via an expectation-conditional maximization algorithm paired with a cross-validation  procedure.
Through a Monte Carlo simulation study, the SaS-Funclust method is shown to outperform  other methods already appeared in the literature, both in terms of clustering performance and interpretability.
Finally,  three real-data examples are presented to demonstrate the favourable performance of the  proposed method.
The SaS-Funclust method is implemented in the  \textsf{R} package \textsf{sasfunclust}, available online at \url{https://github.com/unina-sfere/sasfunclust}.
\end{abstract}

\noindent%
\vfill

\newpage
\spacingset{1.5} 
\section{Introduction}
\label{sec:intro}
In the last years, due to advances in technology and computational power, most of the data collected by practitioners and scientists in many fields bring information about curves or surfaces that are apt to be modelled as functional data, i.e., continuous random functions defined on a compact domain.
A thorough overview of functional data analysis (FDA)  techniques  can be found in \cite{ramsay2005functional,ramsay2009functional,horvath2012inference,hsing2015theoretical} and \cite{kokoszka2017introduction}.
As in the classical (non-functional) statistical literature, cluster analysis is an important
topic in FDA, with many applications in various fields. The primary concern of functional clustering techniques is to classify a sample of data into homogeneous groups of curves, without having any  prior knowledge about the true underlying clustering structure.
The clustering of functional data is generally a difficult task because of the infinite dimensionality of the problem. For this reason,  methods for functional data clustering  have received a lot of attention in recent years, and different approaches have been proposed and discussed in the last decade.
To the best of authors' knowledge, the most used approach is the filtering approach \citep{jacques2014functional}, which relies on the reduction of the infinite dimensional problem  by approximating functional
data in a finite dimensional space  and, then, uses traditional clustering tools on  the basis expansion coefficients.
Along this line, \cite{abraham2003unsupervised} propose an advanced version of the k-means algorithm to the coefficients obtained by projecting the original profiles onto a lower-dimensional subspace spanned by B-spline basis functions. A similar method is proposed by \cite{rossi2004clustering} who apply a Self-Organizing Map (SOM) on the resulting coefficient  instead of the k-means algorithm. 
Elaborating on this path, \cite{serban2005cats} present a technique for the nonparametric estimation and clustering of a large number of functional data  that is still based on the k-means algorithm applied to the basis expansion coefficients obtained through  smoothing techniques.  
A step forward is moved by  \cite{chiou2007functional}, who introduce the k-centers functional clustering method to  account, differently from the previous methods, for both the means and the mode of variation differentials between clusters by predicting cluster membership with a reclassification step.

Instead  of considering the basis expansion coefficients as parameters, a different idea is that of using  a model-based approach where  coefficients are treated as random variables themselves with a cluster-specific probability distribution.
The seminal work of \cite{james2003clustering} is the first one to develop a flexible model-based procedure to cluster functional data based on a random effects model for the coefficients. This allows for borrowing strength across curves and, thus, for superior results when 
data contain a large number of sparsely sampled curves.
More recently, \cite{bouveyron2011model} propose  a  new functional clustering method, which is referred to as funHDDC and  based on a functional latent Gaussian  mixture model, to fit the functional data in group-specific functional subspaces. By constraining model parameters within and between groups, they obtain a family of parsimonious models that allow for more flexibility.
Analogously, \cite{jacques2013funclust} assume  cluster-specific Gaussian distribution on the principal components resulting from the Karhunen–Loeve expansion of the curves, and \cite{giacofci2013wavelet} propose to use a Gaussian mixture model on the wavelet
decomposition of the curves,  which turns out to be particularly appropriate for peak-like data, as opposed to methods based on splines.

In the multivariate cluster analysis, some attributes could be, however, completely noninformative for uncovering  the clustering structure of interest. As an example, this  often happens in  high-dimensional problems, i.e., where the number of variables is considerably larger than the number of observations. In this setting, the task of  identifying  the features  in which respect true clusters  differ the most is of great interest to achieve  \textit{(a)} a more accurate identification of the groups,   as noninformative features may hide the true clustering structure, and \textit{(b)} an higher interpretability of the analysis, by imputing the presence of the clustering structure to a small number of features.
More in general, the methods capable of selecting informative features and eliminating noninformative ones are referred to as \textit{sparse}.
Such class of methods can be  usually reconducted and regarded as  variable selection methods.
Sparse clustering has received increasing attention in the recent literature.
Based on conventional heuristic clustering algorithms, \cite{friedman2004clustering} develop a new procedure to automatically  detect  subgroups of objects, which preferentially cluster on subsets of features.
 \cite{witten2010framework} elaborate a novel clustering framework based on an adaptively chosen subset of  features that are selected by means of a lasso-type penalty.
In terms of model-based approaches, the method  introduced by \cite{raftery2006variable}  is able to sequentially compare  nested models through approximate Bayes factor and to select the informative features.
\cite{maugis2009variable} improve this method by considering the noninformative features as independent from some informative ones.

It is moreover worth  mentioning quite promising variable selection approaches that make use of  a regularization framework.
The seminal work in this direction is that of \cite{pan2007penalized}, who introduce a penalized likelihood approach with
an $L_1$ penalty function, which is able to automatically achieve variable selection via thresholding and delivering a sparse solution. Similarly, \cite{wang2008variable} suggest a solution by replacing the $L_1$ penalty with either the $L_\infty$ penalty 	or the hierarchical penalization function, which  take into account the fact that cluster means, corresponding to the same feature, can be treated as grouped. \cite{xie2008variable} also account for  grouped parameters through the use of two planes of grouping, named vertical and horizontal grouping.
In all   sparse clustering methods just mentioned, a feature is selected if it is informative for at least one pair of clusters and eliminated otherwise, i.e., if it is noninformative for all clusters. However, some variables could be informative only for specific pairs of clusters. For this reason, \cite{guo2010pairwise} propose a pairwise fusion penalty that penalizes, for each feature, the differences between all pairs of cluster means  and  fuses only the non separated  clusters.

Only recently, the notion of sparseness has been translated into a functional data clustering framework. Specifically, sparse functional clustering methods  aim to cluster the curves while jointly detecting the most informative portion of domain to the clustering in order  to improve both the accuracy and the interpretability of the analysis.
 Based on  the idea of \cite{chen2014optimally},  \cite{floriello2017sparse} propose a sparse functional clustering method based on the estimation of a suitable weight function that is capable of identifying the informative part of the domain. \cite{vitelli2019novel} proposes a novel framework for sparse functional clustering that also embeds an alignment step.
To the best of the authors' knowledge, these are the only works that propose sparse functional clustering methods so far.

In this article, we present  a model-based procedure for the sparse clustering of functional data, named  sparse and smooth functional clustering (SaS-Funclust) method, where the basic idea is to provide accurate and interpretable cluster analysis.
Specifically, the parameters of a general functional Gaussian mixture
model  are estimated by maximizing a penalized version of the log-likelihood function, where  a  functional adaptive pairwise fusion penalty, the functional extension of the penalty proposed by \cite{guo2010pairwise}, is introduced.
Firstly, it penalizes the pointwise differences between all pairs of cluster functional means and locally shrinks the means of  cluster pairs to some common values.
Secondly,  a roughness penalty on  cluster functional means is  considered to further improve the interpretability of the cluster analysis.
Therefore, the SaS-Funclust method gains the ability to detect, for each cluster pair, the informative portion of domain to the clustering, hereinafter always intended in terms of mean differences. If a specific mean pair is fused over a portion of the domain, it is labelled as noninformative to the clustering of that pair. Otherwise, it is labelled as informative. 
In other words, the proposed method is able to detect portions of domain that are  noninformative \textit{pairwise}, i.e., for at least a specific cluster pair, differently from the  method proposed by \cite{floriello2017sparse} that is only able to detect portions of domain that are noninformative \textit{overall}, i.e., for all the cluster pairs simultaneously. Moreover, the model-based fashion of the proposed method provides greater  flexibility than the latter, which basically relies on k-means clustering.
A specific expectation-conditional maximization (ECM) algorithm is designed to perform the maximization of the penalized log-likelihood function, which is a non-trivial problem, and a cross-validation based procedure is  proposed to select the appropriate model.
The  method presented in this article is implemented in the  \textsf{R} package \textsf{sasfunclust}, openly available online at \url{https://github.com/unina-sfere/sasfunclust}.

\begin{figure}
	
	\centering
	\begin{subfigure}[b]{0.3\textwidth}
		
		\centering
		
		\includegraphics[width=\textwidth]{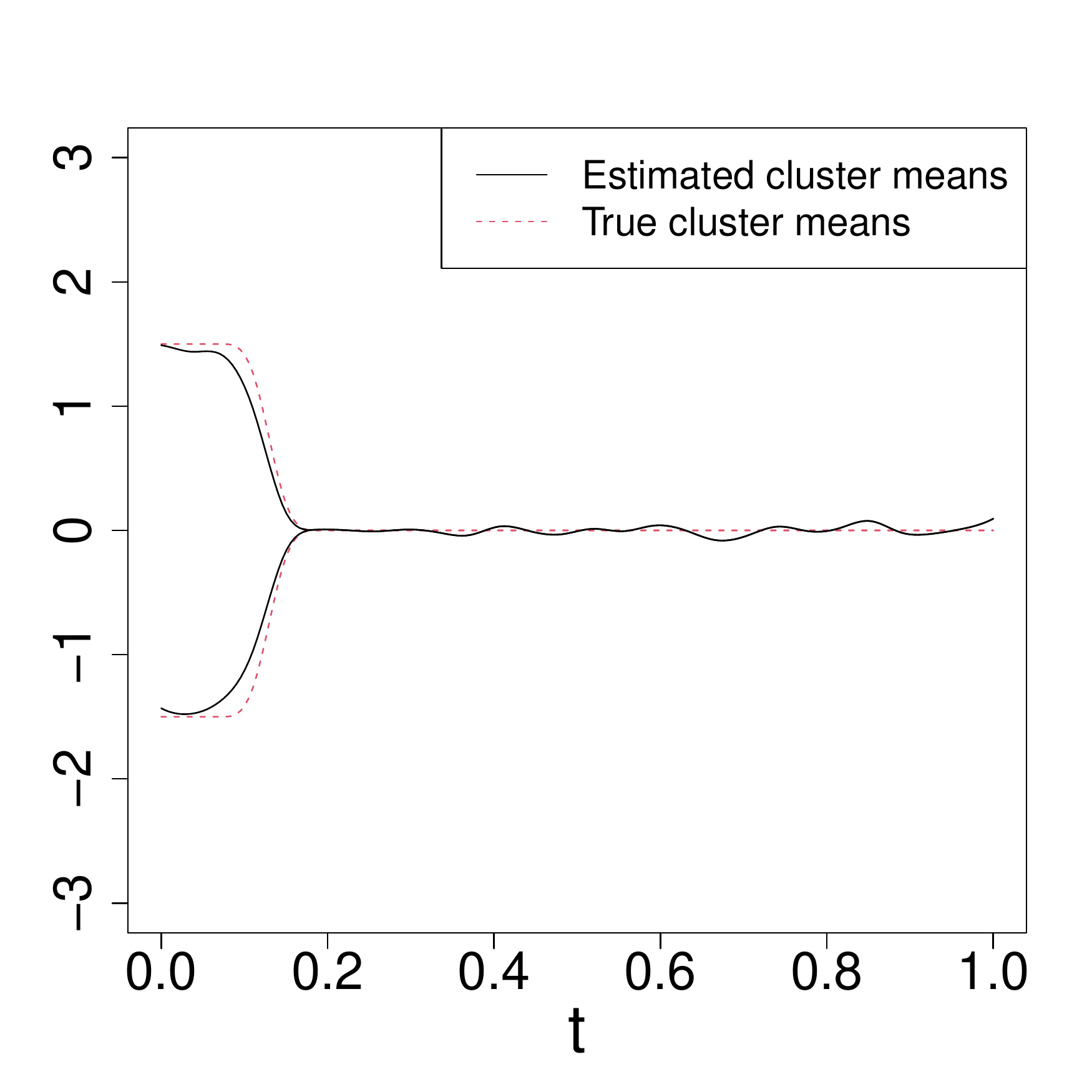}
		\vspace{-1.2cm}
		\caption{}
	\end{subfigure}
	\centering
	\begin{subfigure}[b]{0.3\textwidth}
		
		\centering
		\includegraphics[width=\textwidth]{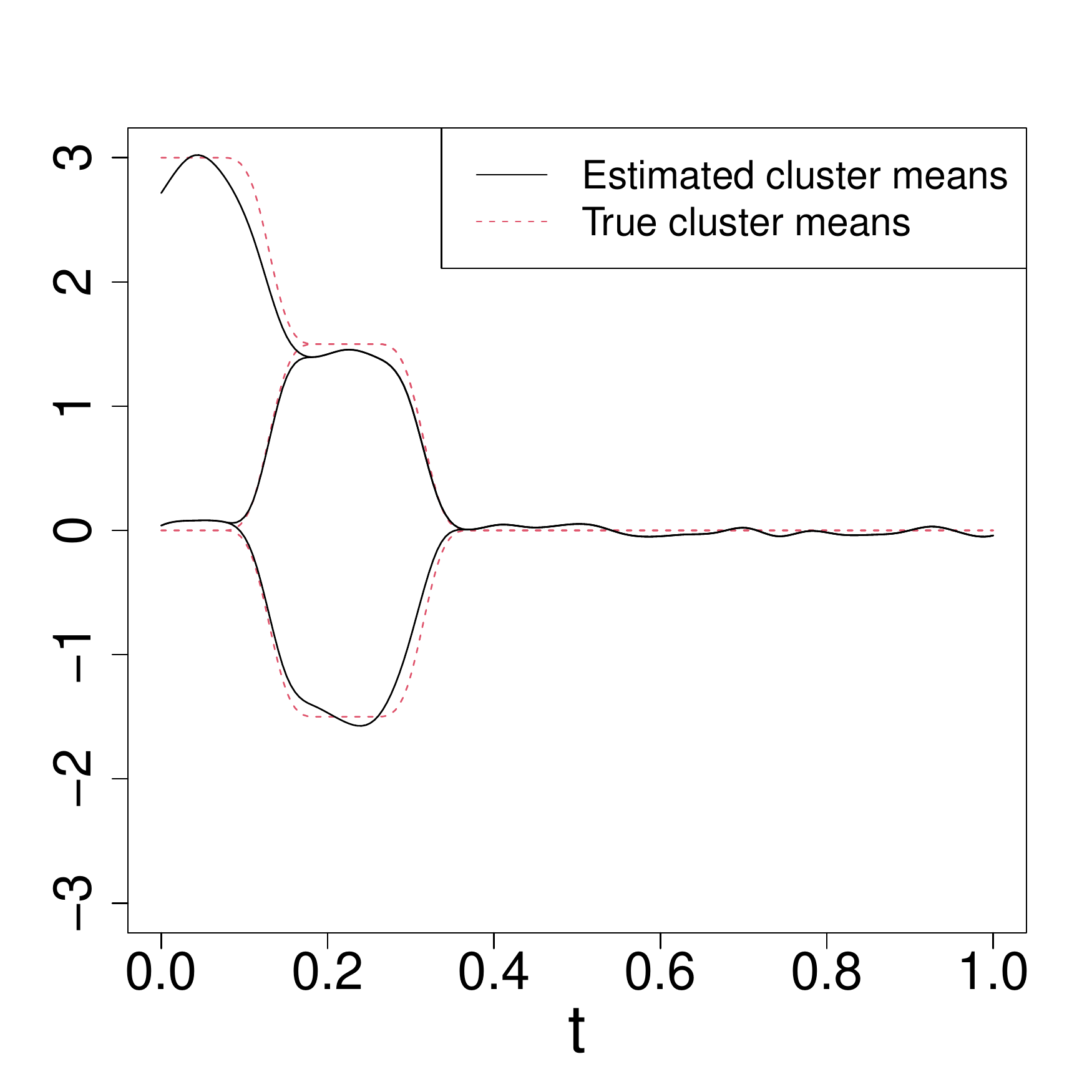}
		\vspace{-1.2cm}
		\caption{}
	\end{subfigure}
	\begin{subfigure}[b]{0.3\textwidth}
		
		\centering
		\includegraphics[width=\textwidth]{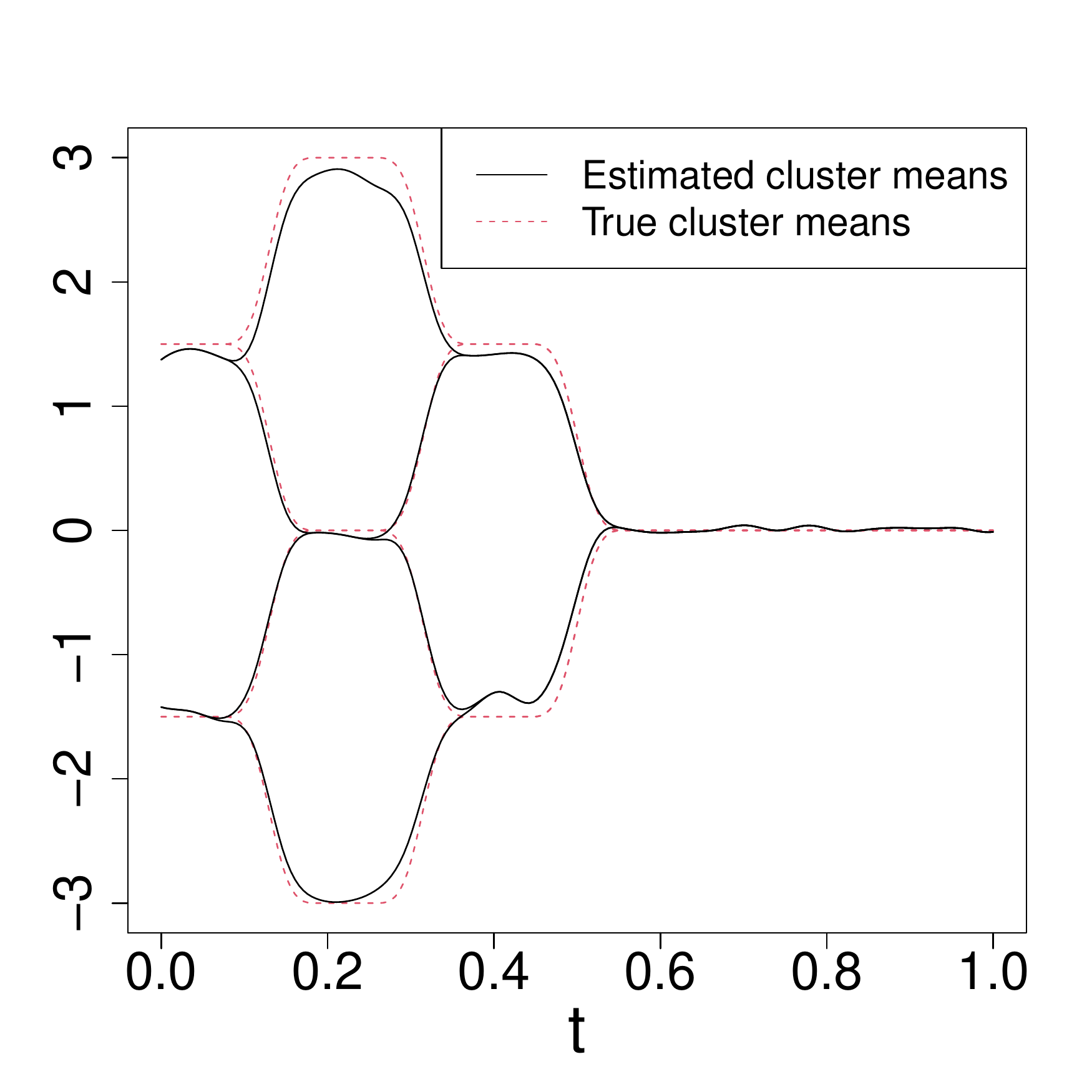}
		\vspace{-1.2cm}
		\caption{}
	\end{subfigure}
	\caption{True and estimated cluster means obtained through the SaS-Funclust method for three different simulated data sets with (a) two, (b) three  and (c) four  clusters generated as described in Section \ref{sec_sim}.}
	\label{fig_mean}
	
\end{figure}

To give a general  idea of the sparseness property of the proposed method, Figure \ref{fig_mean} shows  the cluster means estimated by the latter for three different simulated data sets with (a) two, (b) three and (c) four clusters. Data are generated as described in Section \ref{sec_sim}.
In Figure \ref{fig_mean}(a), the estimated means are correctly fused over $ t\in (0.2,1.0]$. Hence, the proposed method is shown to be able to identify the informative  portion of domain $ [0.0,0.2] $, for the unique pair of clusters and not for all.
In Figure \ref{fig_mean}(b) and Figure \ref{fig_mean}(c), several cluster pairs are available, because the number of clusters is larger than 2, and, thus, a given portion of domain could be informative  for a specific pair of clusters.
 In Figure \ref{fig_mean}(b), the informative portion of domain for each pair of clusters is correctly recovered. The estimated cluster means are indeed pairwise fused over approximately the same portion of domain as the true cluster means pairs.
 Note that, for the clusters whose true means are equal over $ t\in (0.2,1.0]$, the SaS-Funclust method identifies the informative portion of domain roughly in $ [0.0,0.2] $.
 In Figure \ref{fig_mean}(c), the sparseness property of the SaS-Funclust method  is even more striking. In this case, in the face of many cluster pairs,  the proposed method is still able to successfully detect the informative portion of domain.
 The properties of the proposed method will be deepened in Section \ref{sec_sim}.
 
The  remainder of  this  article  is  organized  as  follows. 
Section \ref{sec_met} presents the proposed methodology. Specifically,  Section \ref{sec_genmod} and \ref{sec_pairw} introduce  the general functional Gaussian mixture
model and the penalized maximum likelihood estimator, respectively. Whereas, the optimization algorithm and  model selection considerations are discussed  in Section \ref{sec_opt} and Section  \ref{sec_modsel}, respectively.
Properties and performance of the SaS-Funclust method are assessed through a wide Monte Carlo simulation study presented in Section \ref{sec_sim}.
 Section \ref{sec_rea} illustrates the potentiality of the SaS-Funclust method by means of of  three real datasets: the Berkeley Growth Study, the Canadian weather, and the ICOSAF project data.

\section{The SaS-Funclust method for functional clustering}
\label{sec_met}
\subsection{A general functional clustering model}
\label{sec_genmod}
The SaS-Funclust method is based on the general functional clustering model introduced by \cite{james2003clustering}.
Suppose that $N$ observations are spread among $G$ unknown clusters, and that the probability of each observation to belong to the $g$th cluster is $\pi_g$, $ \sum_{g=1}^{G}\pi_g=1 $.
Moreover, let us denote with $\bm{Z}_i=\left(Z_{1i},\dots,Z_{Gi}\right)^T$ the unknown component-label vector corresponding to the $i$th observation, where $Z_{gi}$ equals 1 if  the $i$th observation is in the $g$th cluster and 0 otherwise.
Then, let us assume that for each $i$ observation, $ i=1,\dots,N $ in the  cluster $ g=1,\dots,G $, it is available a vector $\bm{Y}_i=\left(y_{i1},\dots,y_{in_{i}}\right)^T$ of size $ n_i $, which can differ across observations, of observed values of a function $g_i$  over the time points $t_{i1},\dots,t_{in_{i}}$. The function  $g_i$ is assumed a Gaussian random process with mean $\mu_g$, covariance $\omega_g$,  and values  in $L^2\left(\mathcal{T}\right)$, the separable Hilbert space of square integrable functions defined on the compact domain $\mathcal{T}$.
We assume that, conditionally on that $Z_{gi}=1$, $\bm{Y}_i$ is modelled as
\begin{equation}
	\bm{Y}_i=\bm{g}_i+\bm{\epsilon}_i,\quad i=1,\dots,N,
\end{equation}
where $\bm{g}_i=\left(g_i\left(t_{i1}\right),\dots,g_i\left(t_{in_{i}}\right)\right)^T$ contains the values of the function $g_i$ at $t_{i1},\dots,t_{in_{i}}$ and $\bm{\epsilon}_i$ is a vector of measurement errors that are mutually independent and normally distributed with  mean 0 and    constant variance $\sigma_{e}^2$.
Let us suppose also that the unknown component-label vector $\bm{Z}_i$ has a multinomial distribution, which consists of one draw on $g$ categories with probabilities $\pi_1,\dots,\pi_G$. Then, for every $i$, the unconditional density function $f\left(\cdot\right)$ of 	$\bm{Y}_i$ is
\begin{equation}
\label{eq_mixmod}
f\left(\bm{Y}_i\right)=\sum_{g=1}^{G}\pi_g \psi\left(\bm{Y}_i;\bm{\mu}_{gi},\bm{\Omega}_{gi}+\bm{I}\sigma_{e}^2\right),
\end{equation}
where  $\bm{\mu}_{gi}=\left(\mu_g\left(t_{i1}\right),\dots,\mu_g\left(t_{in_{i}}\right)\right)^T$, $\bm{\Omega}_{gi}=\lbrace \omega_g\left(t_{ki},t_{li}\right)\rbrace_{k,l=1\dots,n_{i}}$, $ \bm{I} $ is the identity matrix, and $\psi\left(\cdot;\bm{\mu},\bm{\Sigma}\right)$ is the multivariate Gaussian density distribution with mean $\bm{\mu}$ and covariance $\bm{\Sigma}$.
The model in Equation \eqref{eq_mixmod} is the classical $G$-component Gaussian mixture model \citep{mclachlan2004finite}.

As discussed in \cite{james2003clustering}, it is necessary to impose some structure curves $g_i$ because both the curves could be observed at different time domain points and the dimensionality of the model in Equation \eqref{eq_mixmod} could be too high in comparison to the sample size.
Therefore, similarly to the filtering approach for clustering, we assume that each function $g_i$, for $ i=1,\dots,N$, may be represented in terms of a $q$-dimensional set of basis functions $\bm{\Phi}=\left(\phi_1,\dots,\phi_q\right)^T$, that is
\begin{equation}
\label{eq_appg}
	g_i\left(t\right)=\bm{\eta}_i^T\bm{\Phi}\left(t\right),\quad t\in\mathcal{T},
\end{equation}
where $ \bm{\eta}_i=\left(\eta_{i1},\dots,\eta_{iq}\right)^T $ are  vectors of basis coefficients.
Then, $ \bm{\eta}_i $ are modelled as Gaussian random vectors, that is, given that  $Z_{gi}=1$,
\begin{equation}
\label{eq_gammai}
	\bm{\eta}_i=\bm{\mu}_g+\bm{\gamma}_{ig},
\end{equation}
where $\bm{\mu}_g=\left(\mu_{g1},\dots,\mu_{gq}\right)^T$ are  $q$-dimensional vectors and $\bm{\gamma}_{ig}$ are  Gaussian random vectors with zero mean and covariance $\bm{\Gamma}_{g}$.

With these assumption the   unconditional density function $f\left(\cdot\right)$ of 	$\bm{Y}_i$ in Equation \eqref{eq_mixmod} becomes
\begin{equation}
f\left(\bm{Y}_i\right)=\sum_{g=1}^{G}\pi_g \psi\left(\bm{Y}_i;\bm{S}_i\bm{\mu}_{g},\bm{\Sigma}_{ig}\right),
\end{equation}
where $\bm{S}_i=\left(\bm{\Phi}\left(t_{i1}\right),\dots,\bm{\Phi}\left(t_{in_{i}}\right)\right)^T$ is the basis matrix for the $i$th curve and $\bm{\Sigma}_{ig}=\bm{S}_i\bm{\Gamma}_g\bm{S}_i^T+\bm{I}\sigma_{e}^2$.
Therefore, the log-likelihood function corresponding to $\bm{Y}_1,\dots,\bm{Y}_N$ is given by
\begin{equation}
\label{eq_loglik}
	L\left(\bm{\Theta}|\bm{Y}_1,\dots,\bm{Y}_N\right)=\sum_{i=1}^{N}\log\sum_{g=1}^{G}\pi_g \psi\left(\bm{Y}_i;\bm{S}_i\bm{\mu}_{g},\bm{\Sigma}_{ig}\right),
\end{equation}
where $\bm{\Theta}=\lbrace \pi_g,\bm{\mu}_g,\bm{\Gamma}_{g},\sigma_{e}^2\rbrace_{g=1,\dots,G}$ is the parameter set of interest.
Based on an estimate $\hat{\bm{\Theta}}=\lbrace \hat{\pi}_g,\hat{\bm{\mu}}_g,\hat{\bm{\Gamma}}_{g},\hat{\sigma}_{e}^2\rbrace_{g=1,\dots,G}$, an observation $ \bm{Y}^{*} $ is assigned to the cluster $ g $ that achieves the largest posterior probability estimate $ \hat{\pi}_g \psi\left(\bm{Y}^{*};\bm{S}_i\hat{\bm{\mu}}_{g},\hat{\bm{\Sigma}}_{ig}\right) $, with $ \hat{\bm{\Sigma}}_{ig}=\bm{S}_i\hat{\bm{\Gamma}}_g\bm{S}_i^T+\bm{I}\hat{\sigma}_{e}^2 $.

\subsection{The penalized maximum likelihood estimator}
\label{sec_pairw}
\cite{james2003clustering} propose to   estimate $\bm{\Theta}$  through the maximum likelihood estimator (MLE), which is the maximizer of the log-likelihood function in Equation \eqref{eq_loglik}. 
In this work, we propose a different estimator $\hat{\bm{\Theta}}_{PMLE}$ of $\bm{\Theta}$  that is the maximizer of the following penalized log-likelihood
\begin{equation}
\label{eq_opt}
	L_p\left(\bm{\Theta}|\bm{Y}_1,\dots,\bm{Y}_N\right)=\sum_{i=1}^{N}\log\sum_{g=1}^{G}\pi_g \psi\left(\bm{Y}_i;\bm{S}_i\bm{\mu}_{g},\bm{\Sigma}_{ig}\right)-\mathcal{P}\left(\bm{\mu}_{g}\right),
\end{equation}
where $\mathcal{P}\left(\cdot\right)$ is a penalty function defined as
\begin{equation}
\label{eq_pen}
\mathcal{P}\left(\bm{\mu}_{g}\right)=\lambda_L\sum_{1\leq g\leq g'\leq G}\int_{\mathcal{T}}\tau_{g,g'}\left(t\right)|\mu_{g}\left(t\right)-\mu_{g'}\left(t\right)|dt+\lambda_s\sum_{g=1}^{G}\int_{\mathcal{T}}\left(\mu_{g}^{(s)}\left(t\right)\right)^2dt,
\end{equation}
where $\lambda_L,\lambda_s\geq 0$ are tuning parameters, and  $\tau_{g,g'}$ are prespecified weight functions. The symbol $f^{(s)}\left(\cdot\right)$ denotes the $s$th-order derivative of  $f$ if the latter a function or the entries of $f$ if it is a vector.
 Note that in Equation \eqref{eq_pen} each function $g_i$ may be represented as in Equation \eqref{eq_appg}, then it follows that
 \begin{equation}
 \label{eq_pen1p}
 \mathcal{P}\left(\bm{\mu}_{g}\right)=\lambda_L\sum_{1\leq g\leq g'\leq G}\int_{\mathcal{T}}\tau_{g,g'}\left(t\right)|\bm{\mu}_{g}^T\bm{\Phi}\left(t\right)-\bm{\mu}_{g'}^T\bm{\Phi}\left(t\right)|dt+\lambda_s\sum_{g=1}^{G}\int_{\mathcal{T}}\left(\bm{\mu}_{g}^T\bm{\Phi}^{(s)}\left(t\right)\right)^2dt,
 \end{equation}
The first element of the right-hand side of  Equation \eqref{eq_pen} is the functional extension of the penalty introduced by \cite{guo2010pairwise} and is referred to as functional adaptive pairwise fusion penalty (FAPFP). The aim of the FAPFP is to shrink the differences between every pair of cluster means for each value of $t\in\mathcal{T}$. Due to the property of the absolute value function of being singular at zero,  some of these differences are shrunken exactly to zero. In particular, the FAPFP allows pair of cluster means to be equal over specific portion of domain that are, thus,  noninformative for separating the means of that pair of clusters.

The choice of the weight function $\tau_{g,g'}$ in Equation \eqref{eq_pen} and Equation \eqref{eq_pen} should be based on the idea that if a given portion of domain is informative for separating the means of the corresponding pair of clusters, then, the values of $\tau_{g,g'}$ over that portion should be small.  In this way, the absolute  difference $ |\mu_{g}\left(\cdot\right)-\mu_{g'}\left(\cdot\right)| $ is  penalized more over noninforvative portions of domain than over informative ones.
Following the standard practice for the adaptive penalties \citep{zou2006adaptive}, we propose to use 
\begin{equation}
\label{eq_weight}
	\tau_{g,g'}\left(t\right)=|\tilde{\mu}_{g}\left(t\right)-\tilde{\mu}_{g'}\left(t\right)|^{-1} \quad t\in\mathcal{T},
\end{equation}
where $\tilde{\mu}_{g}$ are initial estimates of the cluster means.

Finally, the term $ \lambda_s\sum_{g=1}^{G}\int_{\mathcal{T}}\left(\mu_{g}^{(s)}\left(t\right)\right)^2dt $ is a smoothness penalty that penalizes the $s$th derivative of the  cluster means. This term aims to further improve the interpretability of the results by constraining, of a magnitude quantified by $ \lambda_s $, the cluster means  to own a certain degree of smoothness, measured by the derivative order $s$. Following the common practice in FDA \citep{ramsay2005functional}, we suggest  to  penalize  the cluster mean curvature by setting  $s=2$, which implies that the  basis functions chosen are differentiable at least $s$ times.
As a remark, the penalization in Equation \eqref{eq_opt} is applied only to the parameter vectors corresponding to the cluster means, i.e., to $ \bm{\mu}_{1},\dots,\bm{\mu}_{G}$. The reason is that, in this work,  we consider the case where a portion domain is defined as informative   only in terms of   cluster mean differences. However, portions of domain could be informative also in terms of differences in cluster covariances, which together with the means uniquely identify each cluster. 
\subsection{The penalty approximation and the optimization algorithm}
\label{sec_opt}
To perform the maximization of the penalized log-likelihood in Equation \eqref{eq_opt},  the penalty $\mathcal{P}\left(\cdot\right)$, defined as in Equation \eqref{eq_pen}, can be written as 
\begin{align}
\label{eq_pen2}
\mathcal{P}\left(\bm{\mu}_{g}\right)&=\lambda_L\sum_{1\leq g\leq g'\leq G}\int_{\mathcal{T}}|\left(\tilde{\bm{\mu}}_{g}^T-\tilde{\bm{\mu}}_{g'}\right)^T\bm{\Phi}\left(t\right)|^{-1}|\left(\bm{\mu}_{g}^T-\bm{\mu}_{g'}\right)^T\bm{\Phi}\left(t\right)|dt+\lambda_s\sum_{g=1}^{G}\bm{\mu}_{g}^T\bm{W}\bm{\mu}_{g},
\end{align}
where  the  weight functions $ \tau_{g,g'}\left(t\right) $ are expressed as in Equation \eqref{eq_weight},  and  the initial estimates of the cluster means  are  represented  through the set of basis functions $ \bm{\Phi} $ as $\tilde{\mu}_{g}\left(t\right)=\tilde{\bm{\mu}}_{g'}^T\bm{\Phi}\left(t\right)$, $t\in\mathcal T$, with $\tilde{\bm{\mu}}_g=\left(\tilde{\mu}_{g1},\dots,\tilde{\mu}_{gq}\right)^T$. The matrix $\bm{W}$ is equal to $ \int_{\mathcal{T}}\bm{\Phi}^{(s)}\left(t\right)\left(\bm{\Phi}^{(s)}\left(t\right)\right)^Tdt $.
 A great simplification of the optimization problem can be achieved  if the first element on the right-hand side of Equation \eqref{eq_pen2} can be expressed as linear function of $ |\bm{\mu}_{g}^T-\bm{\mu}_{g'}|$. 
 The following theorem provides a practical way to rewrite the first term of the right-hand side of Equation \eqref{eq_pen2} as linear function of $ |\bm{\mu}_{g}^T-\bm{\mu}_{g'}|$, when $\bm{\Phi}$ is a set of B-splines \citep{de1978practical,schumaker2007spline}.
 \begin{theorem}
 	\label{the_1}
 	Let  $\bm{\Phi}=\left(\phi_1,\dots,\phi_{q}\right)^T$ be the set of B-splines of order $k$  and non-decreasing knots sequences $\lbrace x_{0},x_{1},\dots,x_{M_j},x_{M+1}\rbrace$ defined on the compact set $\mathcal{T}=\left[x_{0},x_{M+1}\right]$, with $ q=M+k $, and $ \lbrace \tau_j\rbrace_{j=1}^{q+1} $ being a sequence with  $ \tau_1= x_{0},
 		\tau_j=\tau_{j-1}+\left(x_{\min\left(M+1,j\right)}-x_{\max(0,j-1-k)}\right)/k,\tau_{q+1}=x_{M+1}$. Then,  for each function $f\left(t\right)=\sum_{i=1}^{q}c_i\phi_i\left(t\right)$, $ t\in \mathcal{T} $, where $c_i\in\mathbb{R}$, the function $\tilde{f}\left(t\right)=\sum_{i=1}^{q}c_iI_{\left[\tau_i,\tau_{i+1}\right]}\left(t\right)$, $ t\in \mathcal{T} $, where $ I_{\left[\tau_i,\tau_{i+1}\right]}\left(t\right)=1 $ for $ t\in \left[\tau_i,\tau_{i+1}\right] $ and zero elsewhere, is such that
 		\begin{equation}
 			\sup_{t\in \mathcal{T}}|f\left(t\right)-\tilde{f}\left(t\right)|=O(\delta),
 		\end{equation}
 		where $ \delta=\max_{i}|x_{i+1}-x_{i}| $,
 		that is $f\left(t\right)-\tilde{f}\left(t\right)$ converges uniformly to the zero function.
	
 \end{theorem}
 Theorem \ref{the_1}, whose proof is deferred to the Supplementary material,  basically states that when $ \delta $ is small, $ f $ is well approximated by $ \tilde{f} $. In other words, the approximation error $ |f-\tilde{f}| $ can be made arbitrarily small by increasing  the number of knots. If we further assume  the knots sequence  evenly spaced, $ \delta$ turns out to be  equal to $1/M $.
 These considerations allow us to approximate $ |\left(\bm{\mu}_{g}^T-\bm{\mu}_{g'}\right)^T\bm{\Phi}\left(t\right)| $ and $ |\left(\tilde{\bm{\mu}}_{g}^T-\tilde{\bm{\mu}}_{g'}\right)^T\bm{\Phi}\left(t\right)| $, respectively, as follows
 \begin{equation}\label{eq_app}
 |\left(\bm{\mu}_{g}-\bm{\mu}_{g'}\right)^T\bm{\Phi}\left(t\right)|\approx |\bm{\mu}_{g}-\bm{\mu}_{g'}|^T\bm{I}\left(t\right),\quad |\left(\tilde{\bm{\mu}}_{g}-\tilde{\bm{\mu}}_{g'}\right)^T\bm{\Phi}\left(t\right)|\approx|\tilde{\bm{\mu}}_{g}-\tilde{\bm{\mu}}_{g'}|^T\bm{I}\left(t\right),\quad \forall t \in \mathcal{T},
 \end{equation}
 where $ \bm{I}=\left( I_{\left[\tau_1,\tau_{2}\right]},\dots,I_{\left[\tau_{q},\tau_{q+1}\right]}\right)^T $.
Thus, Equation \eqref{eq_pen2} can be rewritten as
\begin{align}
\label{eq_pen3}
\mathcal{P}\left(\bm{\mu}_{g}\right)&=\lambda_L\sum_{1\leq g\leq g'\leq G}\tilde{\bm{M}}|\bm{\mu}_{g}-\bm{\mu}_{g'}|+\lambda_s\sum_{g=1}^{G}\bm{\mu}_{g}^T\bm{W}\bm{\mu}_{g},
\end{align}
where $ \tilde{\bm{M}}=\diag\left(\frac{\tau_2-\tau_1}{|\tilde{\mu}_{g1}-\tilde{\mu}_{g'1}|},\dots,\frac{\tau_2-\tau_1}{|\tilde{\mu}_{gq}-\tilde{\mu}_{g'q}| }\right) $ is the diagonal matrix with diagonal entries $ \frac{\tau_2-\tau_1}{|\tilde{\mu}_{g1}-\tilde{\mu}_{g'1}|},\dots,\frac{\tau_2-\tau_1}{|\tilde{\mu}_{gq}-\tilde{\mu}_{g'q}| }$.

The goodness of the approximations in Equation \eqref{eq_app} depends on the cardinality $ q $ of the set of B-splines $\bm{\Phi}$, which, thus, should be as large as possible.
However,  the number of parameters  in Equation \eqref{eq_mixmod}, which depends quadratically on $ q $,  becomes very large even for moderate values of $ q $.
To mitigate this issue, one may further   assume equal and diagonal  coefficient covariance matrices  across all clusters, that is $ \bm{\Gamma}_1=\dots=\bm{\Gamma}_G= \bm{\Gamma}=\diag\left(\sigma^2_1,\dots,\sigma^2_q\right)$.
As a remark, with this assumption, we are implicitly assuming that clusters are separated only by their mean values, and, thus, informative portion of domain are identified only by  cluster mean differences and not in terms of covariances.

The penalized log-likelihood function in Equation \eqref{eq_opt} becomes
\begin{equation}
\label{eq_opt_def}
L_p\left(\bm{\Theta}|\bm{Y}_1,\dots,\bm{Y}_N\right)=\sum_{i=1}^{N}\log\sum_{g=1}^{G}\pi_g \psi\left(\bm{Y}_i;\bm{S}_i\bm{\mu}_{g},\bm{\Sigma}_{i}\right)-\lambda_L\sum_{1\leq g\leq g'\leq G}\tilde{\bm{M}}|\bm{\mu}_{g}-\bm{\mu}_{g'}|-\lambda_s\sum_{g=1}^{G}\bm{\mu}_{g}^T\bm{W}\bm{\mu}_{g},
\end{equation}
with $\bm{\Sigma}_{i}=\bm{S}_i\bm{\Gamma}\bm{S}_i^T+\bm{I}\sigma_{e}^2$.
The maximization of this objective function is a nontrival problem. A specifically designed  algorithm is proposed, which is a modification of the expectation maximization (EM) algorithm proposed by \cite{james2003clustering}.
By treating the component-label vectors $ \bm{Z}_i $ (defined at the beginning  of Section \ref{sec_genmod}) and $\bm{\gamma}_{ig}$ in Equation \eqref{eq_gammai} as missing data, the complete penalized log-likelihood is given by 
\begin{align}
\label{eq_opt_def2}
L_{cp}\left(\bm{\Theta}|\bm{Y}_1,\dots,\bm{Y}_N\right)=&\sum_{i=1}^{N}\sum_{g=1}^{G}Z_{gi}\left[\log\pi_g+\log \psi\left(\bm{\gamma}_{ig},0,\bm{\Gamma}\right)+\log \psi\left(\bm{Y}_i;\bm{S}_i\left(\bm{\mu}_{g}+\bm{\gamma}_{ig}\right),\sigma_{e}^2\bm{I}\right)\right]\nonumber\\&-\lambda_L\sum_{1\leq g\leq g'\leq G}\tilde{\bm{M}}|\bm{\mu}_{g}-\bm{\mu}_{g'}|+\lambda_s\sum_{g=1}^{G}\bm{\mu}_{g}^T\bm{W}\bm{\mu}_{g}.
\end{align}
The EM algorithm consists in the  maximization, at each iteration  $ t=0,1,2,\dots $, of the expected value of $ L_{cp} $, calculated with respect the   joint distribution of $ \bm{Z}_i $ and $\bm{\gamma}_{ig}$, given $ \bm{Y}_1,\dots,\bm{Y}_N $ and the current parameter estimates $\hat{\bm{\Theta}}^{\left(t\right)}=\lbrace \hat{\pi}^{\left(t\right)}_g,\hat{\bm{\mu}}^{\left(t\right)}_g,\hat{\bm{\Gamma}}^{\left(t\right)}=\diag\left(\hat{\sigma}^{2\left(t\right)}_1,\dots,\hat{\sigma}^{2\left(t\right)}_q\right),\hat{\sigma}_{e}^{2\left(t\right)}\rbrace_{g=1,\dots,G}$.
The algorithm stops when a pre specified stopping condition is met.
At each $ t $, the expected value of $L_{cp}$ as a function the probability of membership $\pi_1,\dots,\pi_G$ is then maximized by setting
\begin{equation}\label{key}
\hat{\pi}^{\left(t+1\right)}_g=\frac{1}{N}\sum_{i=1}^{N}\hat{\pi}^{\left(t+1\right)}_{g|i},
\end{equation}
with $ \hat{\pi}_{g|i}^{\left(t+1\right)}=\Ex\left(Z_{ig}=1|\bm{Y}_i,\hat{\bm{\Theta}}^{\left(t\right)}\right)=\frac{\hat{\pi}^{\left(t\right)}\psi\left(\bm{Y}_i;\bm{S}_i\hat{\bm{\mu}}^{\left(t\right)}_{g},\hat{\bm{\Sigma}}^{\left(t\right)}_{i}\right) }{\sum_{g'=1}^{G}\hat{\pi}^{\left(t\right)}_{g'}\psi\left(\bm{Y}_i;\bm{S}_i\hat{\bm{\mu}}^{\left(t\right)}_{g},\hat{\bm{\Sigma}}^{\left(t\right)}_{i}\right)}$.
With respect to $ \sigma^2_1,\dots,\sigma^2_q $, $L_{cp}$  is maximized by
\begin{equation}\label{key}
\hat{\sigma}^{2\left(t+1\right)} _j=\frac{1}{N}\sum_{i=1}^{N}\sum_{g=1}^{G}\hat{\pi}^{\left(t+1\right)} _{g|i}\Ex\left(\gamma^2_{ig(j)}|\bm{Y}_i, Z_{gi}=1,\hat{\bm{\Theta}}^{\left(t\right)} \right) \quad j=1,\dots,q,
\end{equation}
where $ \gamma^2_{ig(j)} $ indicates the $j$th entry of $ \bm{\gamma}^2_{ig} $.
The value of $ \Ex\left(\gamma^2_{ig(j)}|\bm{Y}_i, Z_{gi}=1,,\hat{\bm{\Theta}}^{\left(t\right)} \right) $ can be calculated by using the property that the (conditional) distribution of $ \bm{\gamma}_{ig} $ given $ \bm{Y}_i, Z_{gi}=1,\hat{\bm{\Theta}}^{\left(t\right)} $ is Gaussian with mean $ \hat{\bm{\Gamma}}^{\left(t\right)} \bm{S}_i^T\left(\bm{S}_i\hat{\bm{\Gamma}}^{\left(t\right)}\bm{S}_i^T+\bm{I}\hat{\sigma}^{2\left(t\right)}\right)^{-1}\left(\bm{Y}_i-\bm{S}_i\hat{\bm{\mu}}_{g}^{\left(t\right)}\right)$ and covariance  $ \hat{\bm{\Gamma}}^{\left(t\right)}-\hat{\bm{\Gamma}}^{\left(t\right)}\bm{S}_i^T \left(\bm{S}_i\hat{\bm{\Gamma}}^{\left(t\right)}\bm{S}_i^T+\bm{I}\hat{\sigma}^{2\left(t\right)}\right)^{-1}\bm{S}_i\hat{\bm{\Gamma}}^{\left(t\right)}$.
Then, $ \sigma_{e}^2 $ is updated as
\begin{multline}\label{key}
\hat{\sigma}_{e}^{2\left(t+1\right)} =\frac{1}{\sum_{i=1}^{N}n_{i}}\sum_{i=1}^{N}\sum_{g=1}^{G}\left[\hat{\pi}_{g|i}^{\left(t+1\right)} \left(\bm{Y}_i-\bm{S}_i\hat{\bm{\mu}}^{\left(t\right)} _{g}-\bm{S}_i\hat{\bm{\gamma}}_{ig}^{\left(t\right)} \right)\right)^T
\left(\bm{Y}_i-\bm{S}_i\hat{\bm{\mu}}^{\left(t\right)}-\bm{S}_i\hat{\bm{\gamma}}_{ig}^{\left(t\right)}\right)\\
-\bm{S}_i\Cov\left(\bm{\gamma}_{ig}|\bm{Y}_i, Z_{gi}=1,\hat{\bm{\Theta}}^{\left(t\right)} \right)\bm{S}_i^T\left.\right],
\end{multline}
where $ \hat{\bm{\gamma}}_{ig}^{\left(t\right)}=\Ex\left(\bm{\gamma}_{ig}|\bm{Y}_i, Z_{gi}=1,\hat{\bm{\Theta}}^{\left(t\right)} \right) $.

The mean vectors $ \bm{\mu}_1,\dots,\bm{\mu}_{G} $ that maximize the conditional expectation of $L_{cp}$ are the solution of the following optimization problem
\begin{multline}\label{eq_meanmax}
\hat{\bm{\mu}}^{\left(t+1\right)}_1,\dots,\hat{\bm{\mu}}^{\left(t+1\right)}_{G}=\argmin_{\bm{\mu}_1,\dots,\bm{\mu}_{G}}\frac{1}{2}\sum_{i=1}^{N}\sum_{g=1}^{G}\hat{\pi}_{g|i}^{\left(t+1\right)}\frac{1}{\hat{\sigma}_{e}^{\left(t\right)}}\left(\bm{Y}_i-\bm{S}_i\left(\bm{\mu}_{g}+\hat{\bm{\gamma}}_{ig}^{\left(t\right)}\right)\right)^T\left(\bm{Y}_i-\bm{S}_i\left(\bm{\mu}_{g}+\hat{\bm{\gamma}}_{ig}^{\left(t\right)}\right)\right)\\
+\lambda_L\sum_{1\leq g\leq g'\leq G}\tilde{\bm{M}}|\bm{\mu}_{g}-\bm{\mu}_{g'}|+\lambda_s\sum_{g=1}^{G}\bm{\mu}^{T}\bm{W}\bm{\mu}_{g}.
\end{multline}
The optimization problem  in Equation \eqref{eq_meanmax} is a difficult task of the non differentiability of the absolute value function in zero, and, it has not a closed form solution.
However, following the idea of \cite{fan2001variable}, it can be solved by means of the standard local quadratic approximation method, i.e.,  by iteratively solving the following quadratic optimization problem for $ s=0,1,2,\dots $
\begin{multline}\label{eq_meanmax2}
\hat{\bm{\mu}}^{\left(t+1,s+1\right)}_1,\dots,\hat{\bm{\mu}}^{\left(t+1,s+1\right)}_{G}=\argmin_{\bm{\mu}_1,\dots,\bm{\mu}_{G}}\frac{1}{2}\sum_{i=1}^{N}\sum_{g=1}^{G}\hat{\pi}_{g|i}^{\left(t+1\right)}\frac{1}{\hat{\sigma}_{e}^{\left(t\right)}}\left(\bm{Y}_i-\bm{S}_i\left(\bm{\mu}_{g}+\hat{\bm{\gamma}}_{ig}^{\left(t\right)}\right)\right)^T\left(\bm{Y}_i-\bm{S}_i\left(\bm{\mu}_{g}+\hat{\bm{\gamma}}_{ig}^{\left(t\right)}\right)\right)\\
+\lambda_L\sum_{1\leq g\leq g'\leq G}|\bm{\mu}_{g}-\bm{\mu}_{g'}|^T\tilde{\bm{M}}\bm{D}^{\left(s\right)}|\bm{\mu}_{g}-\bm{\mu}_{g'}|+\lambda_s\sum_{g=1}^{G}\bm{\mu}_{g}^{T}\bm{W}\bm{\mu}_{g},
\end{multline}
where $ \bm{D}^{\left(s\right)}=\diag\left(\frac{1}{2|\hat{\mu}^{\left(t+1,s\right)}_{g1}-\hat{\mu}^{\left(t+1,s\right)}_{g'1}|},\dots,\frac{1}{2|\hat{\mu}^{\left(t+1,s\right)}_{gq}-\hat{\mu}^{\left(t+1,s\right)}_{g'q}|}\right) $, and $\hat{\bm{\mu}}^{\left(t+1,0\right)}_1=\hat{\bm{\mu}}^{\left(t\right)}_1,\dots,\hat{\bm{\mu}}^{\left(t+1,0\right)}_G=\hat{\bm{\mu}}^{\left(t\right)}_G$.
Equation \eqref{eq_meanmax2} is based on the following approximation \citep{fan2001variable}
\begin{equation}
\label{eq_meanmax3}
|\mu_{gi}-\mu_{g'i}|\approx\frac{|\mu_{gi}-\mu_{g'i}|}{2|\hat{\mu}^{\left(t+1,s\right)}_{gq}-\hat{\mu}^{\left(t+1,s\right)}_{g'q}|}+\frac{1}{2}|\hat{\mu}^{\left(t+1,s\right)}_{gq}-\hat{\mu}^{\left(t+1,s\right)}_{g'q}|.
\end{equation}
The solution to the original problem in Equation \eqref{eq_meanmax} can be satisfactorily approximated by the solution at iteration $s^{*}$ of the optimization problem in Equation \eqref{eq_meanmax3} when a pre specified stopping condition is met, i.e., $\hat{\bm{\mu}}^{\left(t+1\right)}_1=\hat{\bm{\mu}}^{\left(t+1, s^{*}\right)}_1,\dots,\hat{\bm{\mu}}^{\left(t+1\right)}_G=\hat{\bm{\mu}}^{\left(t+1, s^{*}\right)}_G$.
For numerical stability, a reasonable suggestion is to set a lower bound on $ |\hat{\mu}^{\left(t+1,s\right)}_{gi}-\hat{\mu}^{\left(t+1,s\right)}_{g'i}| $, and then to shrink to zero  all the estimates below the lower bound. 
It is worth noting that the proposed modification to the algorithm of \cite{james2003clustering} falls within the class of the expectation conditional maximization (ECM) algorithms \citep{meng1993maximum}. Based on the convergence property of the ECM algorithms, which also holds for the   local quadratic approximation in variable selection problems \citep{fan2001variable,hunter2005variable}, the proposed algorithm  can be proved to converge to a stationary point of the penalized log-likelihood in Equation \eqref{eq_opt_def}.

\subsection{Model selection}
\label{sec_modsel}
 The proposed SaS-Funclust method requires the choice several hyper-parameters viz., the number of clusters $G$,  tuning parameters $ \lambda_s,\lambda_L $,   dimension $ q $ and the order $ k $ of the set of B-spline functions  $\bm{\Phi}$ as well as  the knot locations. 
A standard choice for $\bm{\Phi}$  is the cubic B-splines (i.e., $ k=4 $) with equally spaced knot sequence,   which enjoy the optimal property of interpolation \citep{de1978practical}.
Moreover, the dimension $ q $ should be set as large as possible  to reduce, to the greatest possible extent, the approximation error in Equation \eqref{eq_app}. This facilitates the estimated cluster means to successfully capture the local feature of the true cluster means. Unfortunately, the larger the value of $ q $, the higher the complexity of the model in Equation \eqref{eq_mixmod}, i.e., the number of parameters to estimate. The presence of the smoothness penalty on  $ \bm{\mu}_g $, as well as the   constraint imposed on  $ \bm{\Gamma}_g $, allows one to control the complexity of the model and, thus, to prevent over-fitting issues.
The choice of $G, \lambda_s$, and $\lambda_L $ may be based on a $ K $-fold cross-validation procedure.
Based on observations  divided into $ K $ equal-sized disjoint subsets $ f_1,\dots,f_k,\dots,f_K $, $G, \lambda_s$, and $\lambda_L $ are chosen as the maximizers of the following function
\begin{equation}\label{eq_cv}
	CV\left(G, \lambda_s,\lambda_L\right)=\frac{1}{K}\sum_{k=1}^{K}\sum_{i\in f_k}\log\sum_{g=1}^{G}\hat{\pi}^{-f_k}_g \psi\left(\bm{Y}_i;\bm{S}_i\hat{\bm{\mu}}^{-f_k}_{g},\hat{\bm{\Sigma}}^{-f_k}_{i}\right),
\end{equation}
where $ \hat{\pi}^{-f_k}_g,\hat{\bm{\mu}}^{-f_k}_{g}$ and $\hat{\bm{\Sigma}}^{-f_k}_{i} $ denote the SaS-Funclust estimates of $ \pi_g,\bm{\mu}_{g}$ and $\bm{\Sigma}_{i} $ obtained by leaving out the observations in the $k$-th subset $f_k $.
Usually,  the $ CV $ function is numerically calculated over a finite grid of values.
As in the multivariate regression setting,  the uncertainty of the $ CV $ function in estimating the  log-likelihood observed for an out-of-sample observation is taken into account by means of the so called  $ m $-standard deviation rule. This heuristic rule suggests to pick up the  most parsimonious model among those achieving values of the $ CV $ function that are  no more than $m$ standard errors below the maximum of Equation \eqref{eq_cv}.
Note that, in this problem,  parsimony is reflected into large  $ \lambda_s,\lambda_L $ and small $ G $. 
By elaborating on the $ m$-standard deviation rule, we propose to firstly choose  
$G $  for each value of $ \lambda_s,\lambda_L $, with $ m=m_1 $; secondly, at fixed $G $, choose   $ \lambda_s$ for each $ \lambda_L $, with $ m=m_2 $;  thirdly, to choose  $ \lambda_L$ at fixed $ \lambda_s$ and $G$, by using $ m=m_3 $.
In this way,  the estimated model is not unnecessarily complex and achieves predictive performance that is comparable to that of  the best model (i.e., the one that maximizes the $ CV $ function in Equation \eqref{eq_cv}).

As a remark, although the component-wise procedure proposed to choose $\lambda_s, \lambda_L$ and $G $ proves itself to be very effective in the simulation study of Section \ref{sec_sim},  we recommand whenever possible to directly plot and inspect  the $ CV $ curve as a function of $G, \lambda_s$, and $\lambda_L $  and to use any information available from the specific application. 

\section{Simulation study}
\label{sec_sim}
In this section, the performance of the SaS-Funclust method  is assessed by means of an extensive Monte	Carlo simulation study. 
The SaS-Funclust method, implemented through the \textsf{R} package \textsf{sasfunclust},  is compared with the following methods that have already appeared in the literature before. In particular, we refer to the method proposed by \cite{giacofci2013wavelet}  as curvclust, and to that proposed by \cite{bouveyron2011model} as funHDDC. These methods are implemented through the homonymous  \textsf{R} packages \textsf{curvclust} \citep{curvclust} and \textsf{funHDDC} \citep{funHDDC}.
 In addition, we consider also filtering approaches based on two main steps. The first step consists in the estimation of the functions $ g_i $ by means of either smoothing B-splines  or functional principal component analysis \citep{ramsay2005functional}; whereas the second step aims to apply  standard clustering algorithms, viz.  hierarchical,  k-means and finite mixture model clustering methods \citep{everitt2011cluster},  on either the  resulting B-spline coefficients or the functional principal components scores.
Filtering approaches  based on the smoothing B-splines and the hierarchical, k-means  and finite mixture model clustering methods will be hereinafter referred to as   B-HC, B-KM and B-FMM, respectively, whereas methods based on the functional principal component analysis and the  hierarchical, k-means  and finite mixture model clustering methods are referred to as FPCA-HC, FPCA-KM and FPCA-FMM.
Finally, we evaluate also the method presented by  \cite{ieva2013multivariate}, which is referred to as DIS-KM and it basically consists in the application of  the k-means clustering to the $ L^2 $ distances among the observed curves.

The number of clusters is selected through the Bayesian information criterion (BIC)  for the curvclust and funHDDC methods, as suggested by \cite{giacofci2013wavelet} and \cite{bouveyron2011model}, respectively; whereas the silhouette index \citep{rousseeuw1987silhouettes} is used for the DIS-KM method. The majority rule applied to several validity indices \citep{charrad2012nbclust}  is used to determine the number of clusters for all the filtering approaches.
The number of clusters and the tuning parameters needed to implement the SaS-Funclust method  are determined through the CV based procedure described in Section \ref{sec_modsel} with $ q=30 $, $ K=5 $, $ m_1=m_3=0.5 $, and $ m_2=0 $. The values of $ m_1$ and $m_3$ ensure parsimony in the choice of $\lambda_L$ and $G$, whereas for picking $\lambda_s$ the  $m$-standard deviation  is not applied.
 The initial values of the parameters for the ECM algorithm are chosen by applying the  k-means algorithm on the  coefficients estimated through smoothing B-spline.

The performance of the clustering procedures in selecting the  proper number of clusters and  identifying the clustering structure, when the true number of cluster is known, is assessed separately.
In particular, the former is measured through the mean number of selected clusters, whereas the latter is compared through the  adjusted Rand index \citep{hubert1985comparing} denoted by $aRand$. This index accounts for the agreement between  true data partitions and  clustering results corrected  by chance, based on the number of paired objects that are either in the same group or in different groups in both partitions. The $aRand$   yields  values between 0 and 1. The larger its value, the higher the similarity between the two partitions.

 Three different scenarios are analysed where data are generated from $ G_t= 2,3,4  $ clusters and referred to as Scenario I, II and III, respectively. 
 For each scenario, the considered methods are evaluated by assessing the performance over 100 independently generated datasets.
 From each cluster,  200 observations are generated over the domain $\mathcal{T}=\left[0,1\right]$.  The  true functions  are obtained as $ g_i=\bm{\eta}_i^{T}\bm{\Theta}_B$, with  $ \bm{\Theta}_{B} $ representing a set of $ 30 $ evenly spaced knot cubic B-splines. The coefficients $ \bm{\eta}_i=\left(\eta_{i1},\dots,\eta_{i30}\right)^T $ are  Gaussian random coefficients with mean vectors (depending on both  the scenario and the cluster  considered) reported in Table \ref{ta_1}, and covariance matrix $ \bm{\Gamma}=\sigma_c^2\bm{I} $, where $ \sigma_c=0.5 $ and $ \bm{I} $ denotes the identity matrix. Then, to obtain the contaminated-with-error observations $ \bm{Y}_i $,  the  true functions $ g_i $ are evaluated over a grid of $ 50 $ points, with the addition of measurement errors independently  generated as Gaussian random variables with zero mean  and five different values of standard error $ \sigma_e= 1,1.5,	2,2.5,3 $.
 Note that, from Scenario I to Scenario III,   the portion of domain that is noninformative for \textit{all} cluster pairs decreases, whereas, the  portions of domain that are informative for \textit{specific} cluster pairs increases.  
 
 \begin{table}
 	\caption{Coefficient mean vectors for each scenario and cluster.\label{ta_1}}
 	\begin{center}
 		\resizebox{0.35\textwidth}{!}{
 			\begin{tabular}{cccccc}
 				\toprule
 				&$ \bm{\eta}_i $&Cluster 1& Cluster 2& Cluster 3&Cluster 4\\
 				\midrule
 				\multirow{2}{*}{Scenario I}&$\eta_{i1},\dots,\eta_{i5}$&1.5&-1.5&-&-\\
 				&$\eta_{i6},\dots,\eta_{i30}$&0&0&-&-\\
 				\midrule
 				\multirow{3}{*}{Scenario II}&$\eta_{i1},\dots,\eta_{i5}$&3&0&0&-\\
 				&$\eta_{i6},\dots,\eta_{i10}$&1.5&1.5&-1.5&-\\
 				&$\eta_{i11},\dots,\eta_{i30}$&0&0&0&-\\
 				\midrule
 				\multirow{4}{*}{Scenario III}&$\eta_{i1},\dots,\eta_{i5}$&1.5&1.5&-1.5&-1.5\\
 				&$\eta_{i6},\dots,\eta_{i10}$&3&0&0&-3\\
 				&$\eta_{i11},\dots,\eta_{i15}$&1.5&1.5&-1.5&-1.5\\
 				&$\eta_{i16},\dots,\eta_{i30}$&0&0&0&0
 				\\
 				\bottomrule
 			\end{tabular}
 		}
 	\end{center}
 	
 \end{table}

\begin{figure}
	
	\centering
	\begin{subfigure}[b]{0.3\textwidth}
		
		\centering
		
		\includegraphics[width=\textwidth]{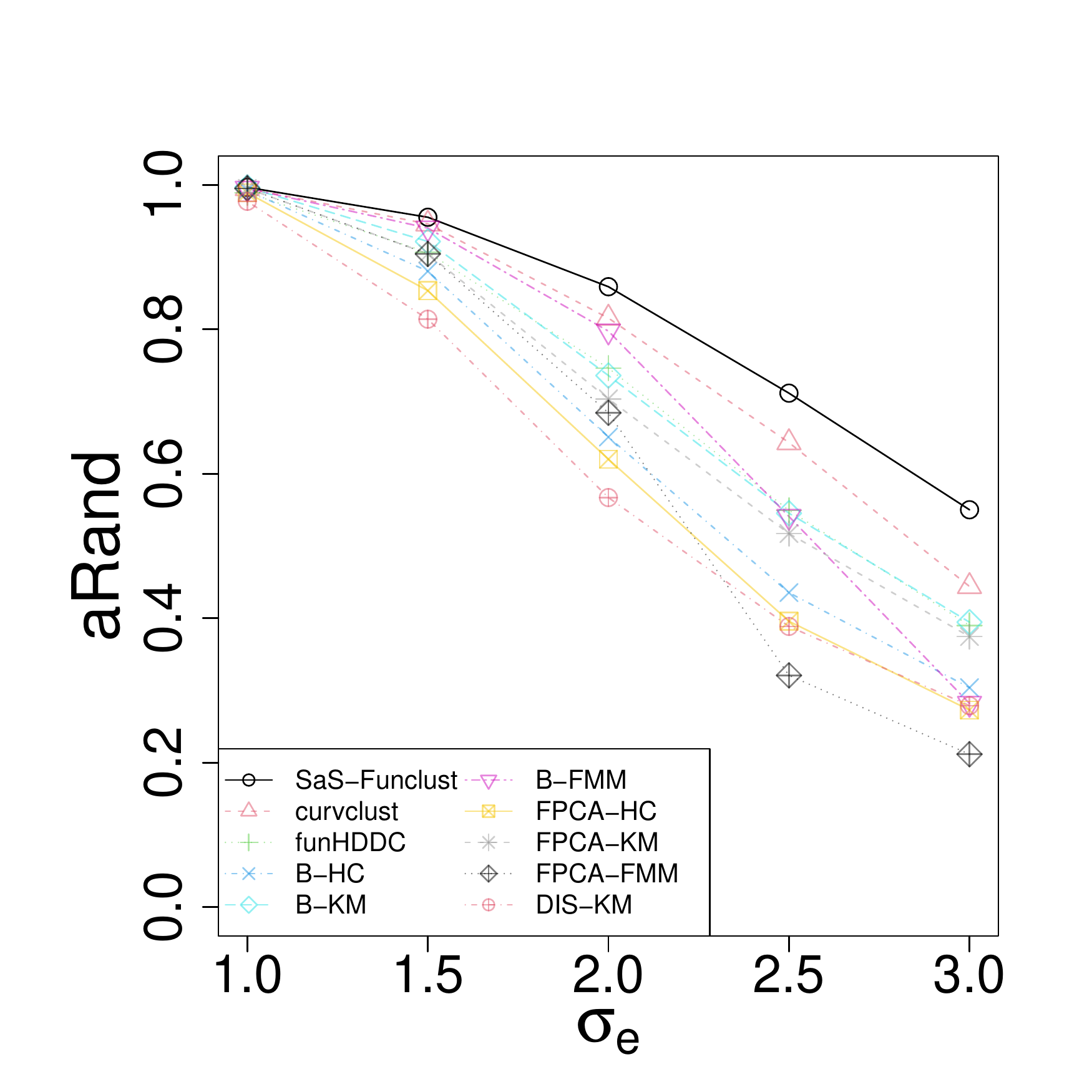}
		\vspace{-1.2cm}
		\caption{}
	\end{subfigure}
	\centering
	\begin{subfigure}[b]{0.3\textwidth}
		
		\centering
				\includegraphics[width=\textwidth]{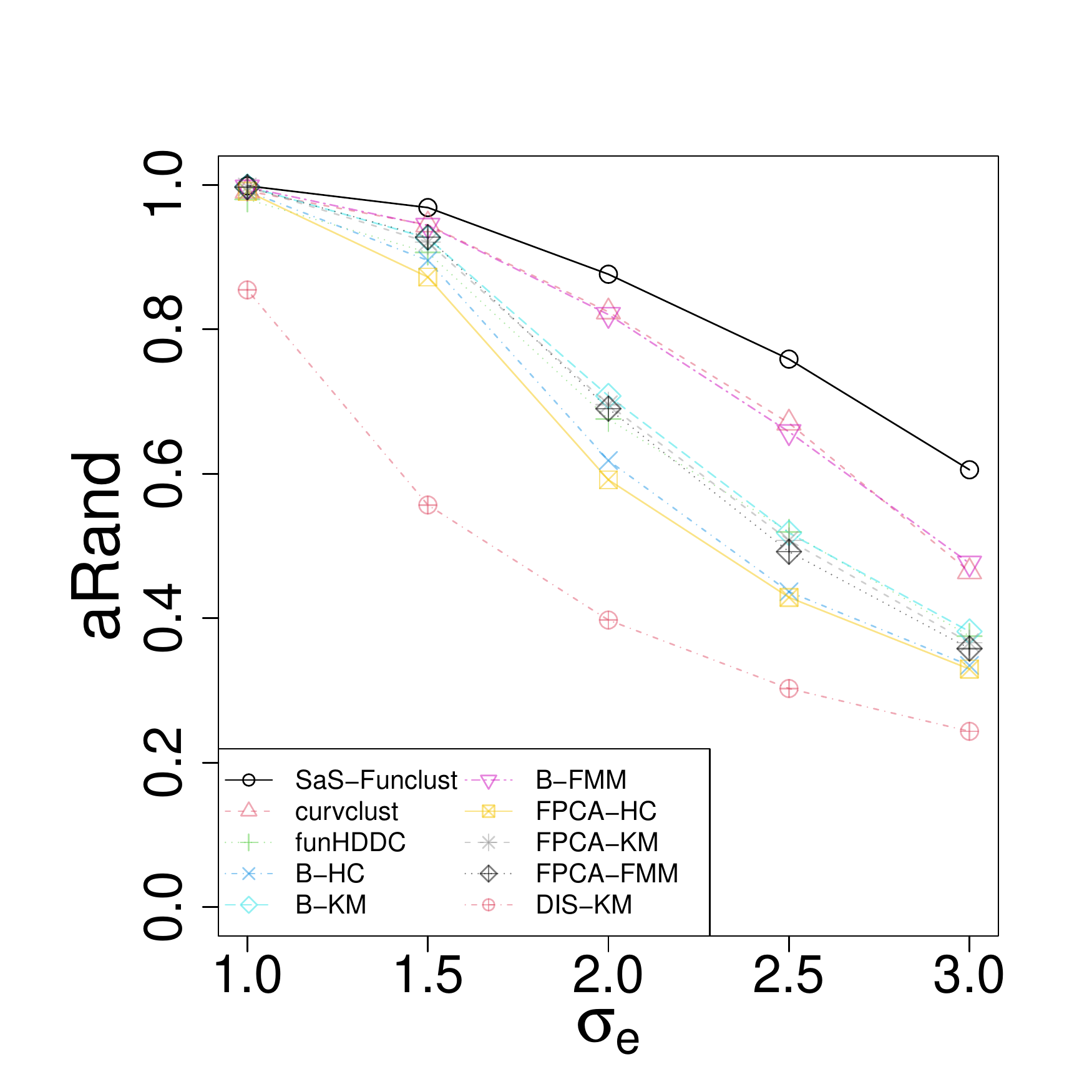}
		\vspace{-1.2cm}
		\caption{}
	\end{subfigure}
\begin{subfigure}[b]{0.3\textwidth}
	
	\centering
	\includegraphics[width=\textwidth]{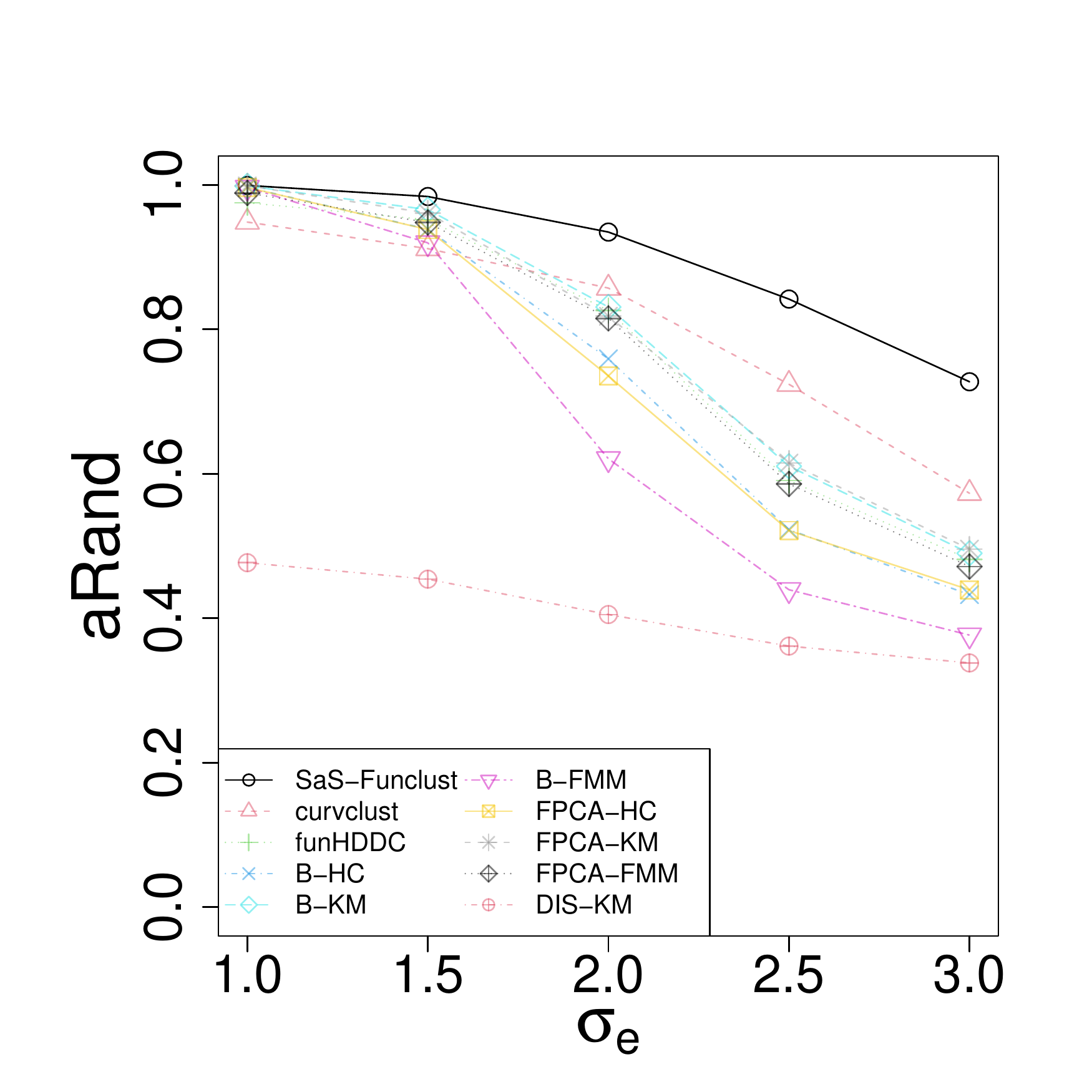}
	\vspace{-1.2cm}
	\caption{}
\end{subfigure}
	\caption{Average $aRand$ index for  (a) Scenario I, (b) Scenario II, and  (c) Scenario III  as a function of $ \sigma_{e} $ when the true number of clusters is known.}
	\label{fig_arand}
	
\end{figure}

Figure \ref{fig_arand} shows the average $aRand$ index values for Scenario I, through III as a function of the standard error $ \sigma_{e} $.
In Scenario I, at small values of $ \sigma_{e} $, all methods perform comparably and provide clustering partitions with $aRand$ values very close to 1, which corresponds to the perfect cluster identification.
However, as $ \sigma_{e} $ increases, the SaS-Funclust method turns out to be the best method,  closely followed by the curvclust method.
Also the B-FMM performs very well, except when $ \sigma_{e}=3.0 $. 
In Scenario II and  III, the  SaS-Funclust method is still the best, followed by the curvclust and B-FMM case in Scenario II and only by the curvclust method in Scenario III.
Note that in these scenarios, the DIS-KM underperforms also in the most favourable cases as a consequence of the lesser capacity of the $ L^2 $ distance to  recover the true clustering structure.

\begin{figure}
	
	\centering
	\begin{subfigure}[b]{0.3\textwidth}
		
		\centering
		
		\includegraphics[width=\textwidth]{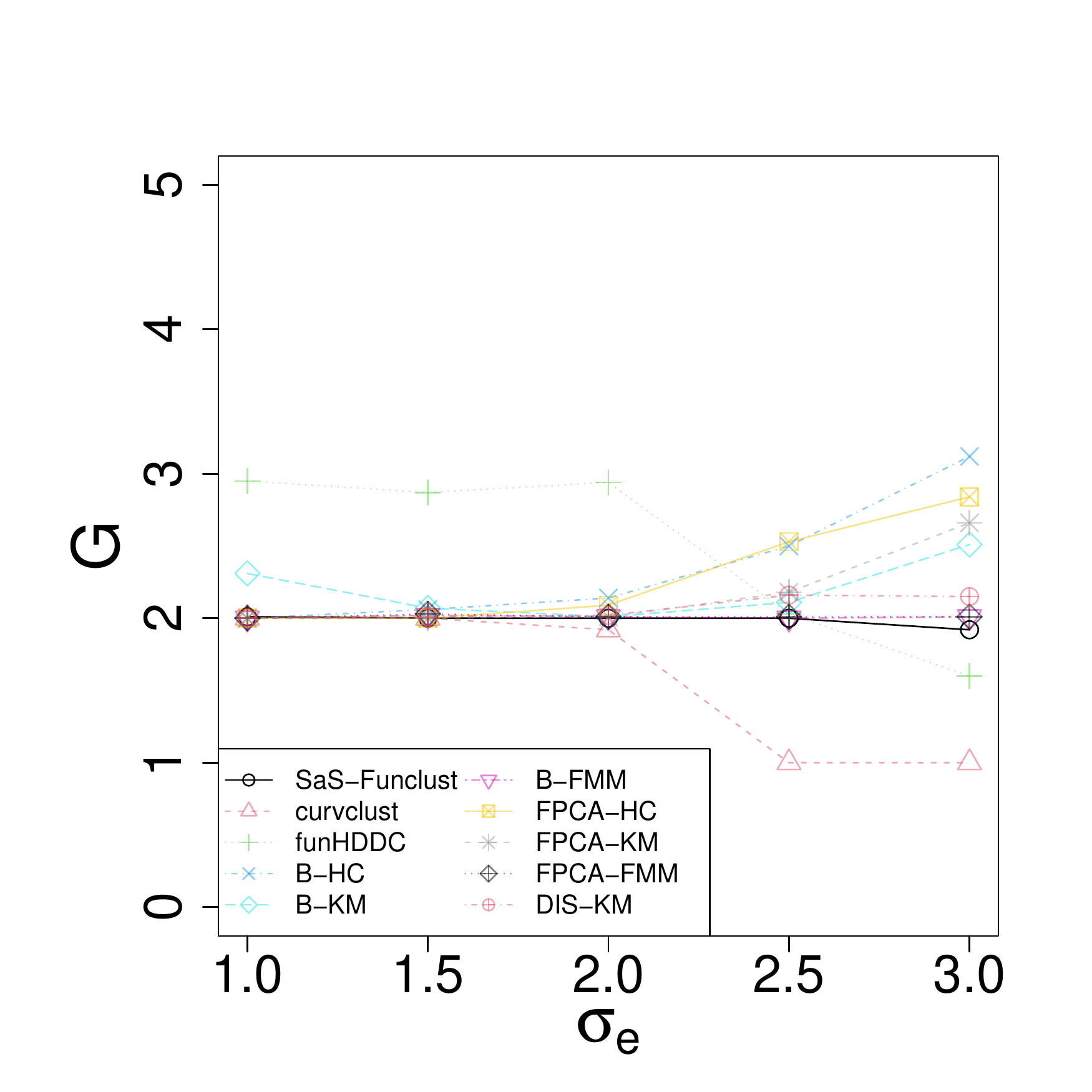}
		\vspace{-1.2cm}
		\caption{}
	\end{subfigure}
	\centering
	\begin{subfigure}[b]{0.3\textwidth}
		
		\centering
		\includegraphics[width=\textwidth]{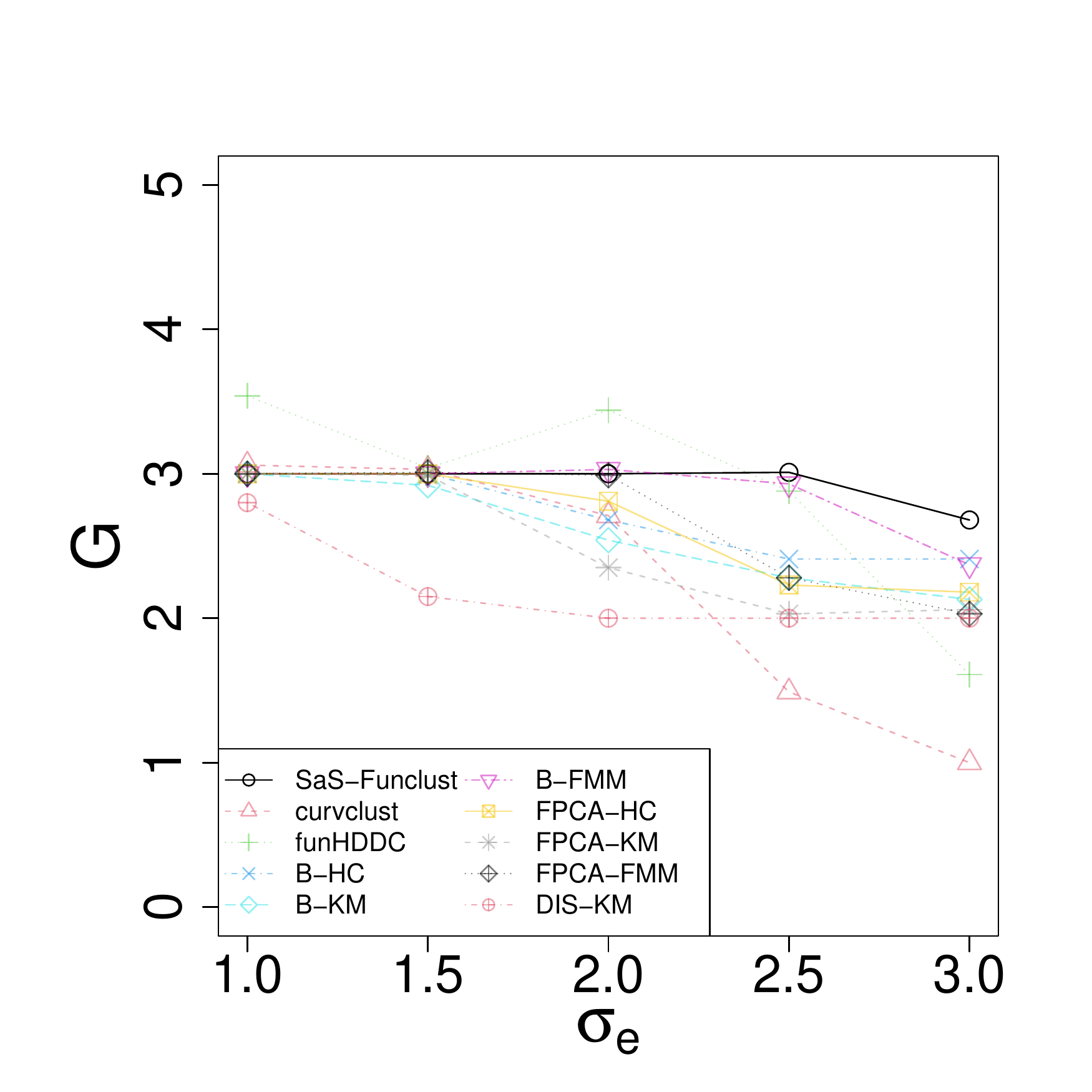}
		\vspace{-1.2cm}
		\caption{}
	\end{subfigure}
	\begin{subfigure}[b]{0.3\textwidth}
		
		\centering
		\includegraphics[width=\textwidth]{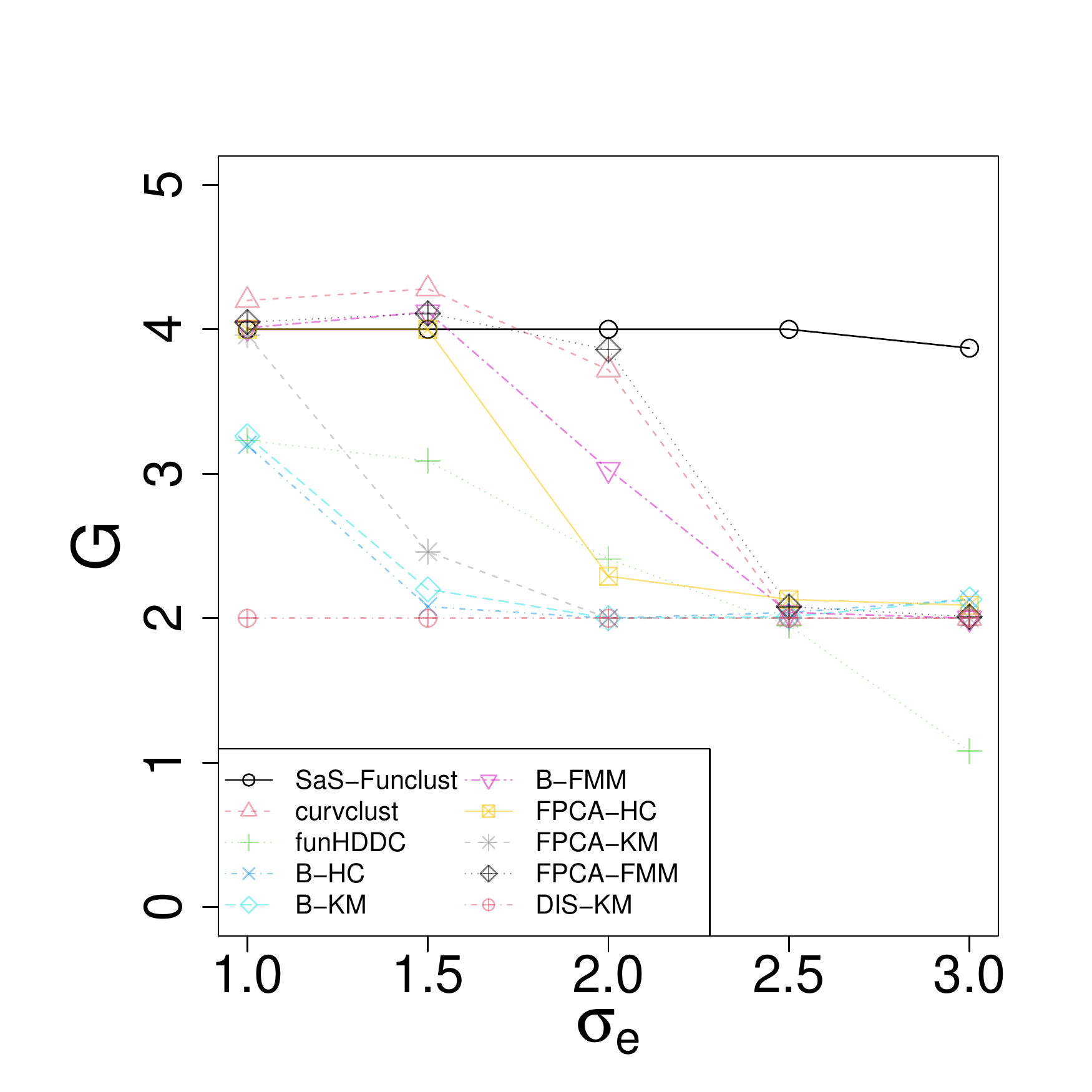}
		\vspace{-1.2cm}
		\caption{}
	\end{subfigure}
	\caption{Average selected number of clusters $G$ for (a) Scenario I, (b) Scenario II, and  (c) Scenario III  as a function of $ \sigma_{e} $.}
	\label{fig_G}
	
\end{figure}
Figure \ref{fig_G} shows the mean number of selected clusters in all  scenarios.
It is  clear that the SaS-Funclust method is able to identify the true number of clusters much better than the competitors in all the  considered scenarios.
In particular,  Scenario II highlights that, especially for large measurement error  $ \sigma_{e} $,  the competing methods reduce  their complexity and select, on average, a number of clusters smaller than the true number of clusters $G_t=3$. 
This is  evident in Scenario III, where the competing methods  select, on average, a number of clusters $G=2$ for $ \sigma_{e}= 2.5,3.0 $, which is smaller than  $G_t=4$. 

\begin{figure}
	
	\centering
	\begin{subfigure}[b]{0.3\textwidth}
		
		\centering
		
		\includegraphics[width=\textwidth]{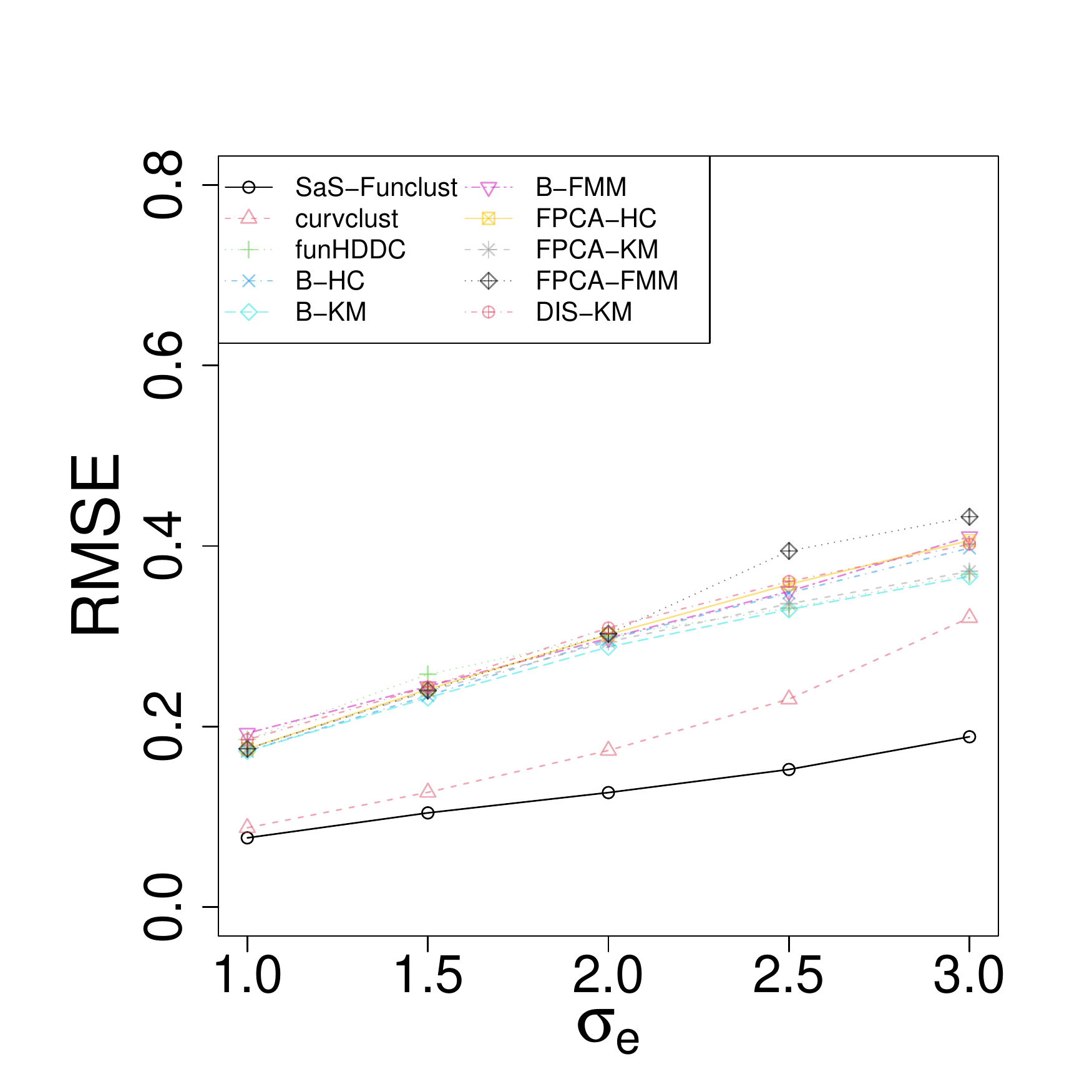}
		\vspace{-1.2cm}
		\caption{}
	\end{subfigure}
	\centering
	\begin{subfigure}[b]{0.3\textwidth}
		
		\centering
		\includegraphics[width=\textwidth]{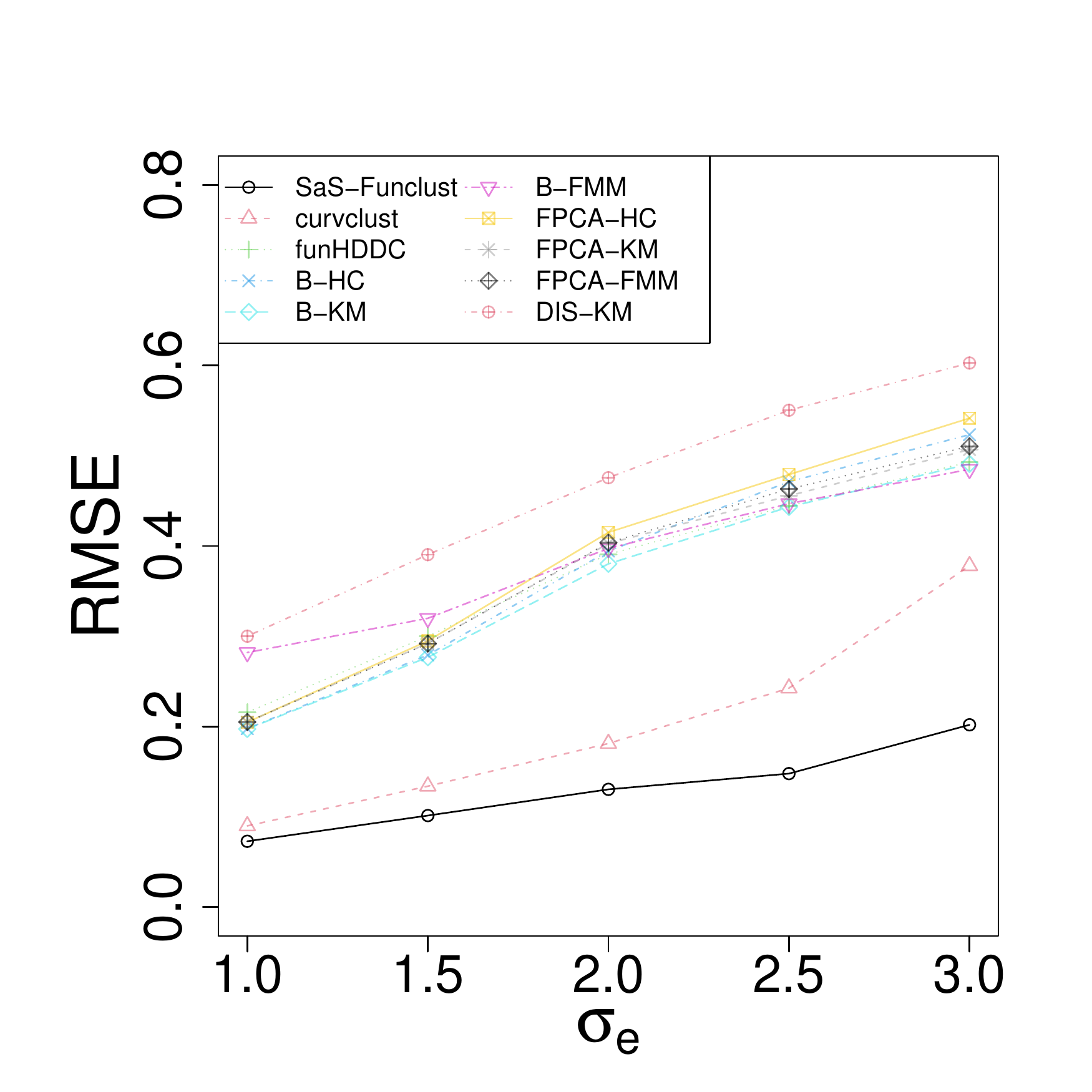}
		\vspace{-1.2cm}
		\caption{}
	\end{subfigure}
	\begin{subfigure}[b]{0.3\textwidth}
		
		\centering
		\includegraphics[width=\textwidth]{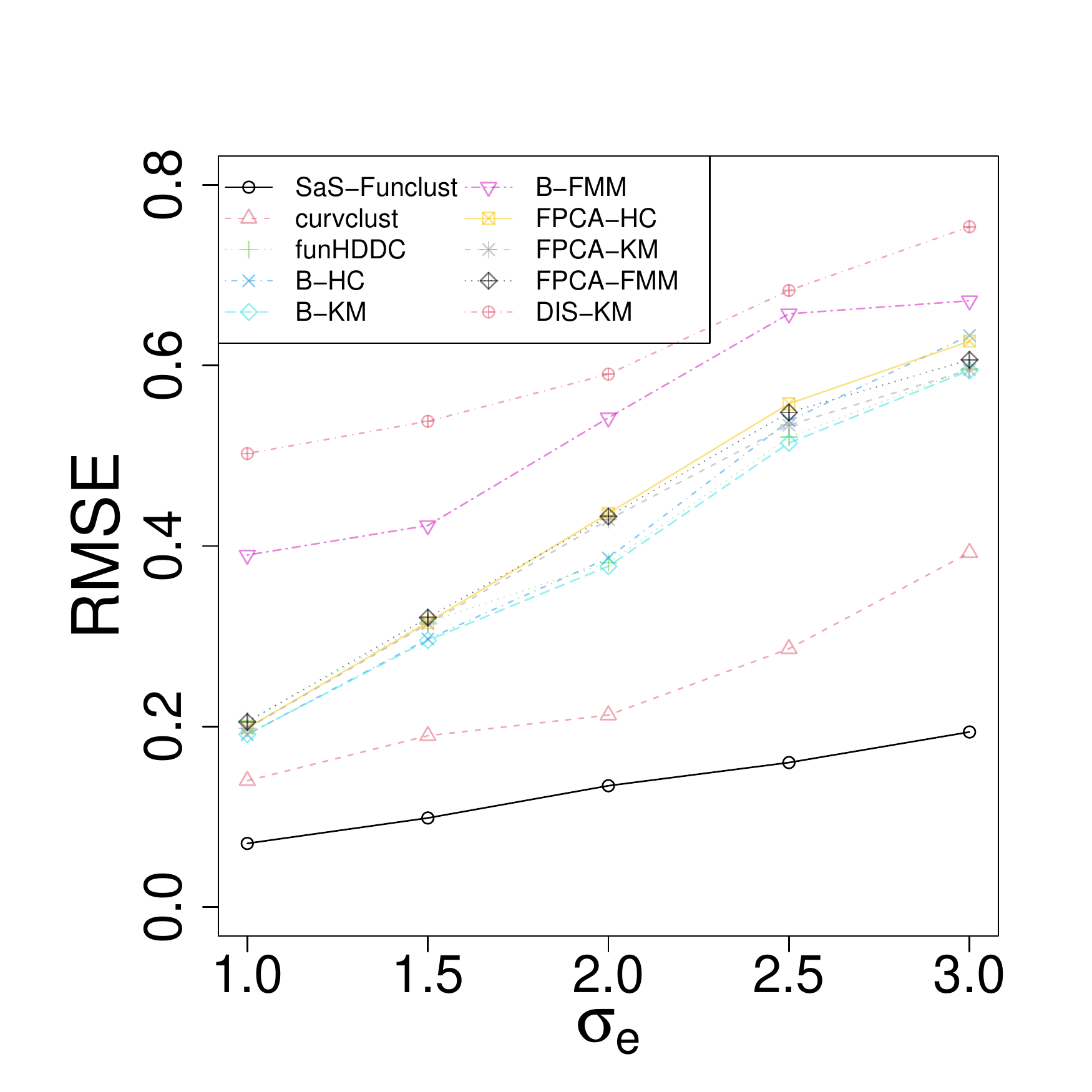}
		\vspace{-1.2cm}
		\caption{}
	\end{subfigure}
	\caption{Average selected number of clusters $G$ for  Scenario I (a),  Scenario II (b), and,   Scenario III (c) as a function of $ \sigma_{e} $.}
	\label{fig_MSE}
	
\end{figure}
Figure \ref{fig_MSE} and Table \ref{tab_zero} highlight the ability of the SaS-Funclust method  in recovering the true  cluster means and  detecting the informative portions of domain.
The average root mean squared error calculated as $ RMSE=\left[\frac{1}{G}\sum_{g=1}^{G}\int_{\mathcal{T}}\left(\mu_g\left(t\right)-\hat{\mu}_g\left(t\right)\right)^2dt\right]^{1/2} $, with $ \hat{\mu}_g\left(t\right)=\hat{\bm{\mu}}_{g}^T\bm{\Phi}\left(t\right)$, $t\in\mathcal T$, is plotted in Figure \ref{fig_MSE} for  each method as a function of $ \sigma_{e} $ in all three scenarios.
By this figure, the SaS-Funclust method outperforms the competitors  in each scenario, especially for large measurement errors, even though the  curvclust method shows comparable performance.
Table \ref{tab_zero} reports, for each  $ \sigma_{e} $ and  scenario, the average fractions of correctly identified noninformative portions of domain by the SaS-Funclust method, which can be regarded as a measure of the interpretability (i.e., sparseness) of the proposed solution.
In more detail, each entry of the table is obtained as the mean of the  average fraction of correctly identified noninformative portions of domain, over the 100 generated datasets,  for each pair of clusters, weighted by the size of the corresponding true noninformative portions of domain. In Scenario I, it trivially coincides with the average, because the true number of clusters is $G_t=2$.
The proposed method is clearly able  to provide an interpretable clustering. The fraction of correctly identified noninformative portions of domain is almost larger than or equal to  $ 0.90 $  for $ \sigma_{e}\leq 2.5$ and decreases to $0.80$  for $ \sigma_{e}=3.5 $.
It is worth noting that when $ \sigma_{e}=1.0 $, the pairs of clusters in each scenario are correctly fused over almost all the noninformative portion of domain in terms of mean differences.
This confirms what is shown in Figure \ref{fig_mean} of Section \ref{sec:intro}.

\begin{table}
	\caption{Average fractions of correctly identified noninformative portions of domain by the SaS-Funclust method  for each scenario.\label{tab_zero}}
	\begin{center}
		\resizebox{0.35\textwidth}{!}{
			\begin{tabular}{ccccc}
				\toprule
				&&Scenario I& Scenario II& Scenario III\\
				\midrule
				\multirow{5}{*}{$ \sigma_{e} $}&1.0&0.9956&0.9901    &0.9782\\
				&1.5& 0.9921   &0.9844&0.9627 \\
				&2.0&0.9846&0.9589&0.9389 \\
				&2.5&0.9565&0.9373&0.8942 \\
				&3.0&0.8821&0.8760& 0.8024\\

				\bottomrule
			\end{tabular}
		}
	\end{center}
	
\end{table}

\section{Real-data Examples}
\label{sec_rea}
\subsection{ Berkeley Growth Study Data	}
\begin{figure}
	\centering
	\includegraphics[width=0.33\textwidth]{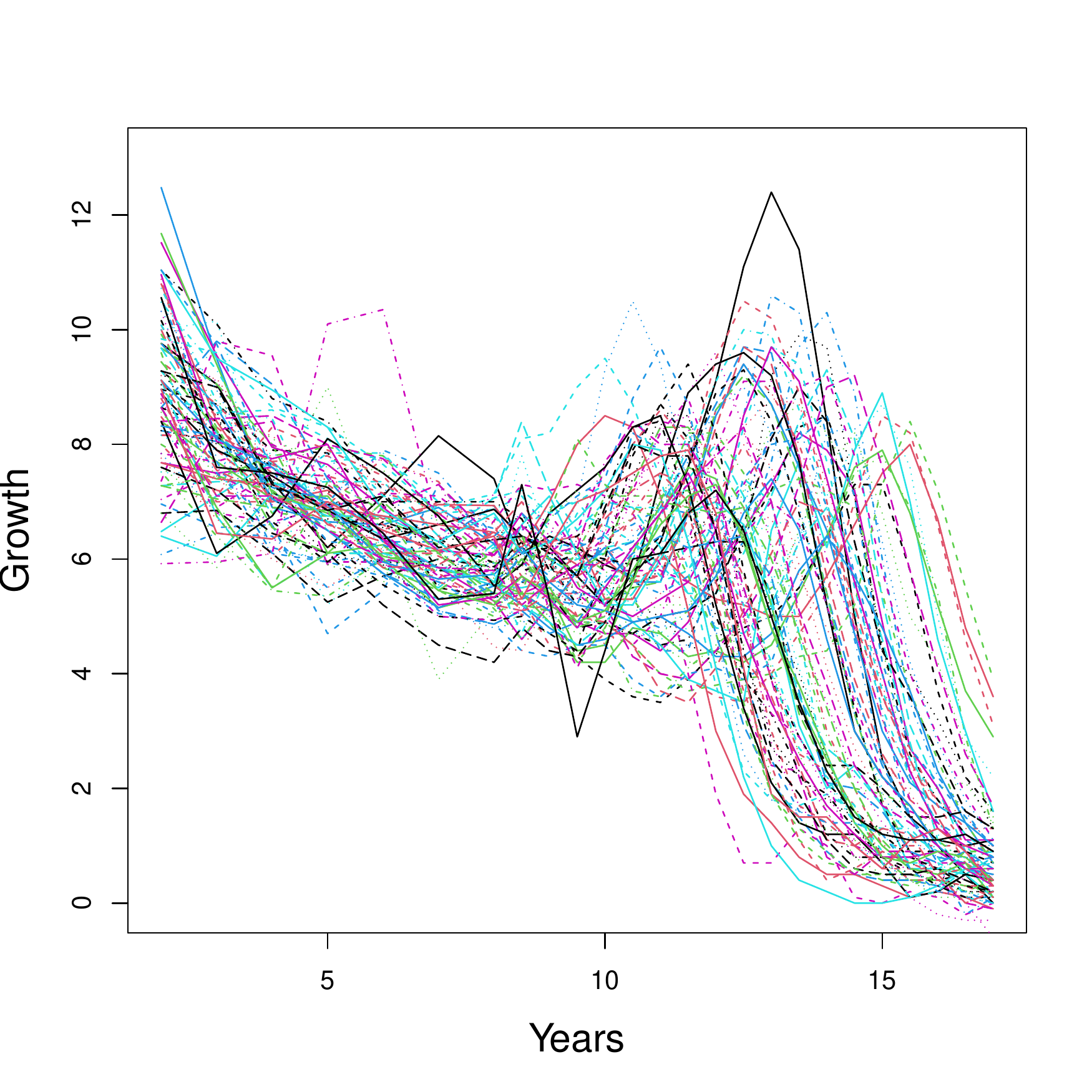}
	\caption{ Growth velocities of 54 girls and 39 boys in the Berkeley growth study dataset.}
	\label{fig_growth}
\end{figure}

In this section, the SaS-Funclust method is applied to the growth dataset from the Berkeley growth study \citep{tuddenham1954physical}, which is available in the \textsf{R} package \textsf{fda} \citep{fda_package}.
In this study, the heights of 54 girls and 39 boys were measured 31 times at age  1 through 18.
The aim of the analysis is to cluster the growth curves and compare the results with the partition based on the gender difference.
This problem has been already addressed  by   \cite{chiou2007functional,jacques2013funclust,floriello2017sparse}.
In particular, we  focus on the growth velocities  from age 2 to 17, whose discrete values are estimated through the central differences method applied to the  growth curves.  Figure \ref{fig_growth} shows the interpolating growth velocity curves for all the individuals.

In view of the  analysis objective,  all  clustering methods described in Section \ref{sec_sim} are applied by setting $ G=2 $.
As shown in the first row of Table \ref{ta_growth}, all  clustering methods, excluded the  B-HC,  perform similarly in terms of the $aRand$ index with respect to the gender difference partition. Moreover, by looking at the second row of Table \ref{ta_growth}, which shows the $aRand$ index with respect to the  SaS-Funclust partition,  the competing methods provide partitions very similar to the SaS-Funclust one.
\begin{table}
	\caption{The values of the $aRand$ index for all the  clustering methods with respect to   gender difference based grouping and the SaS-Funclust partition for the Berkeley growth study dataset \label{ta_growth}}
	\begin{center}
		\resizebox{\textwidth}{!}{
			\begin{tabular}{ccccccccccc}
				\toprule
				&SaS-Funclust& curvclust& funHDDC&B-HC&B-KM&B-FMM&FPCA-KM&FPCA-HC&FPCA-FMM&DIS-KM\\
				\midrule
				Gender difference based grouping&0.58& 0.51& 0.61 &  0.20&0.58&0.58& 0.58& 0.58& 0.58& 0.58\\
				SaS-Funclust&-& 0.83& 0.96&0.37& 1.00 & 1.00& 1.00& 1.00&1.00&1.00\\
				\bottomrule
			\end{tabular}
		}
	\end{center}
	
\end{table}

\begin{figure}
	
	\centering
	\begin{subfigure}[b]{0.33\textwidth}
		
		\centering
		\includegraphics[width=\textwidth]{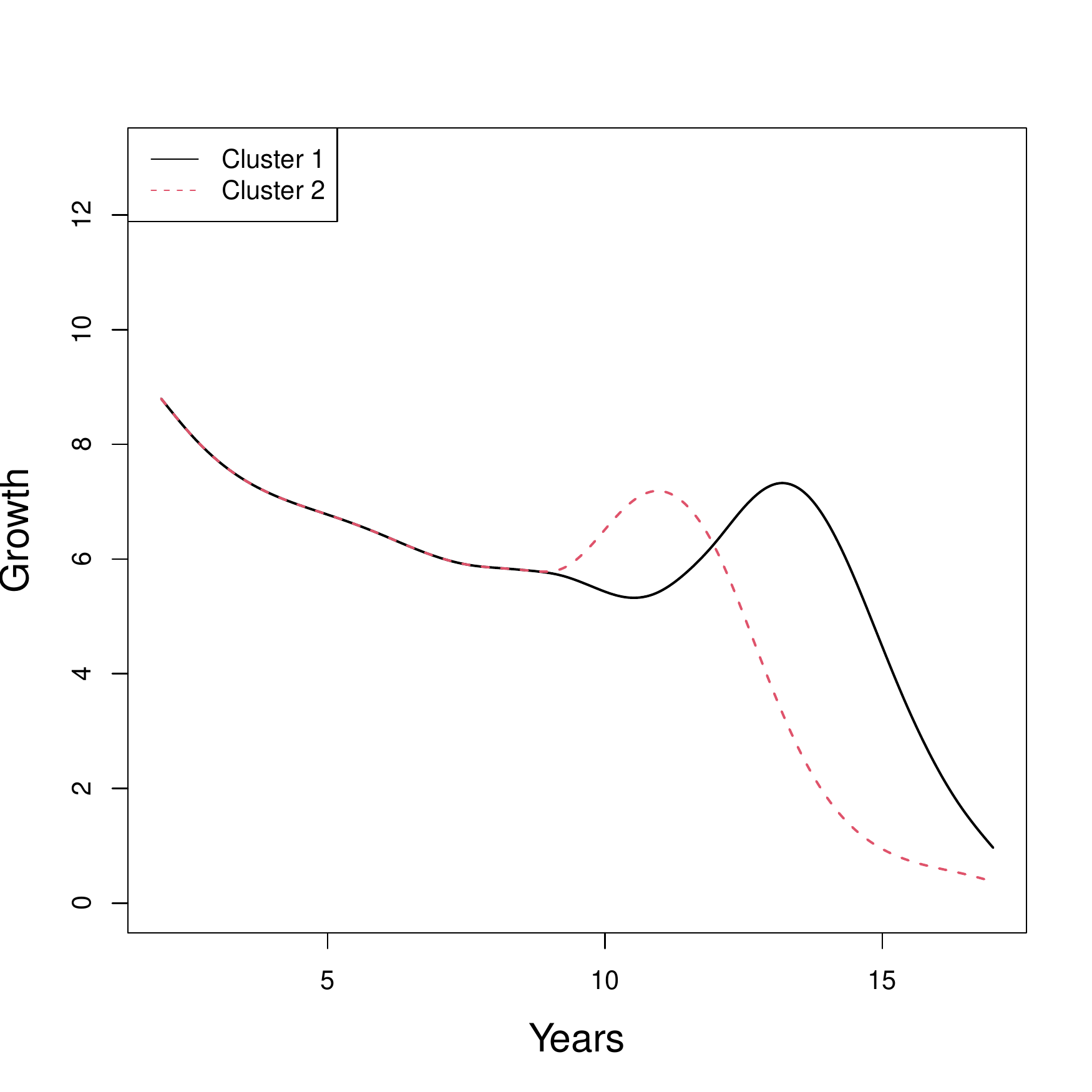}
		\vspace{-1.2cm}
		\caption{}
	\end{subfigure}
	\centering
	\begin{subfigure}[b]{0.33\textwidth}
		
		\centering
		\includegraphics[width=\textwidth]{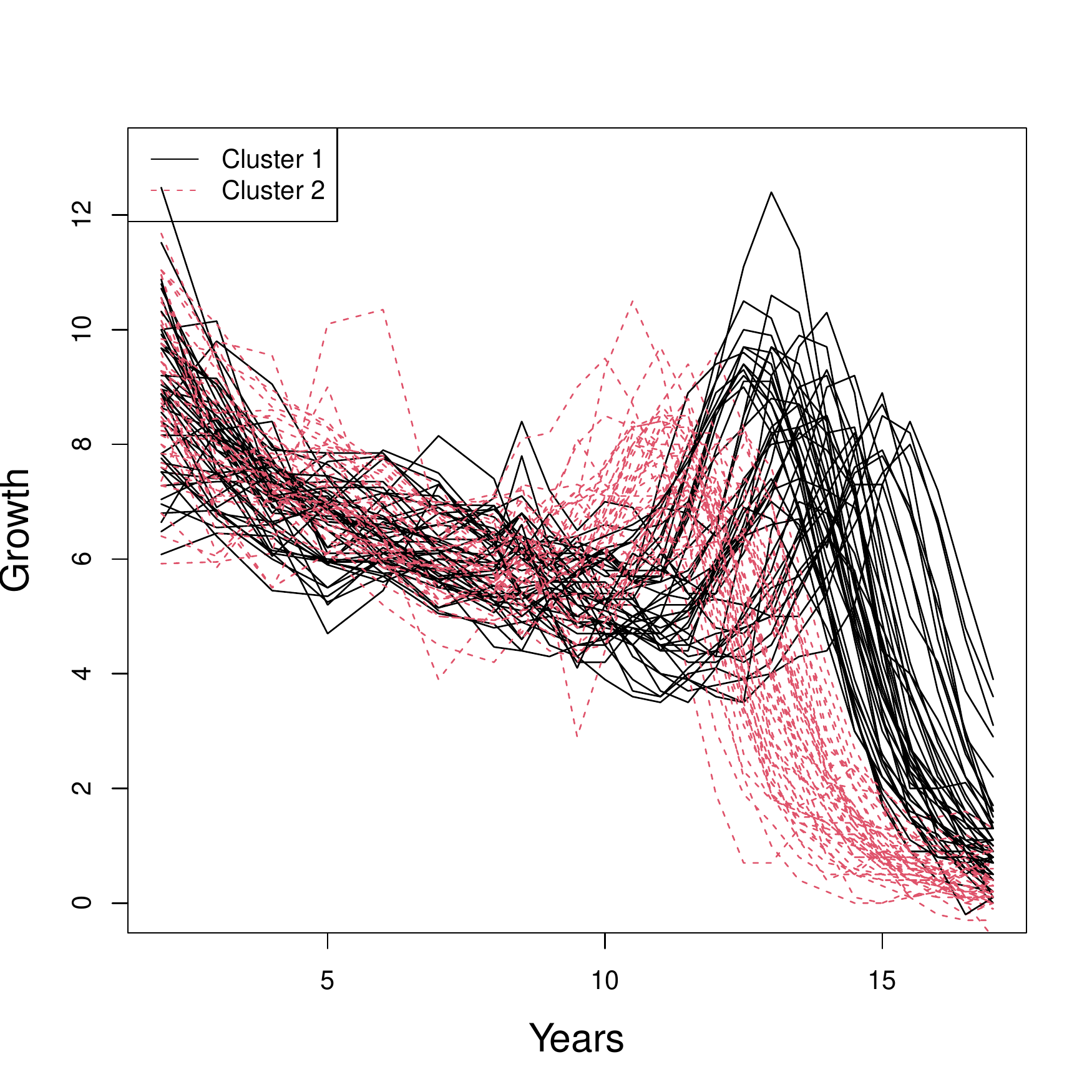}
		\vspace{-1.2cm}
		\caption{}
	\end{subfigure}
	\caption{ (a) Estimated cluster curve means and (b) curve clusters for the SaS-Funclust method  in the Berkeley growth study dataset.}
	\label{fig_growth_mean}
	
\end{figure}

As expected, the SaS-Funclust method  allows for a more interpretable analysis.
Figure \ref{fig_growth_mean} shows the estimated cluster  means and the clustered growth curves for the SaS-Funclust method.
The estimated cluster means are fused over the first portion of the domain, whereas they are separated over the remaining portions.
This implies that the two identified clusters  are not different on average over the first portion of domain which can be, thus,  regarded as noninformative. Separation between the two group arises over the remaining informative portion of domain, where  two sharp peaks of growth velocity arise, instead. 
The latter peaks are known in the medical literature as pubertal spurts, in which respect the attained results  indicate two main timing/duration groups. In particular, male pubertal spurt happens later  and lasts longer than female one. Nevertheless,  some individuals show  unusual growth patterns that are not captured by the cluster analysis.
Additionally,  the estimated cluster means from the competing methods, not shown here, do not allow for a similar straightforward  interpretation.

\subsection{Canadian Weather Data}
\begin{figure}
 	
 	\centering
 	\includegraphics[width=0.33\textwidth]{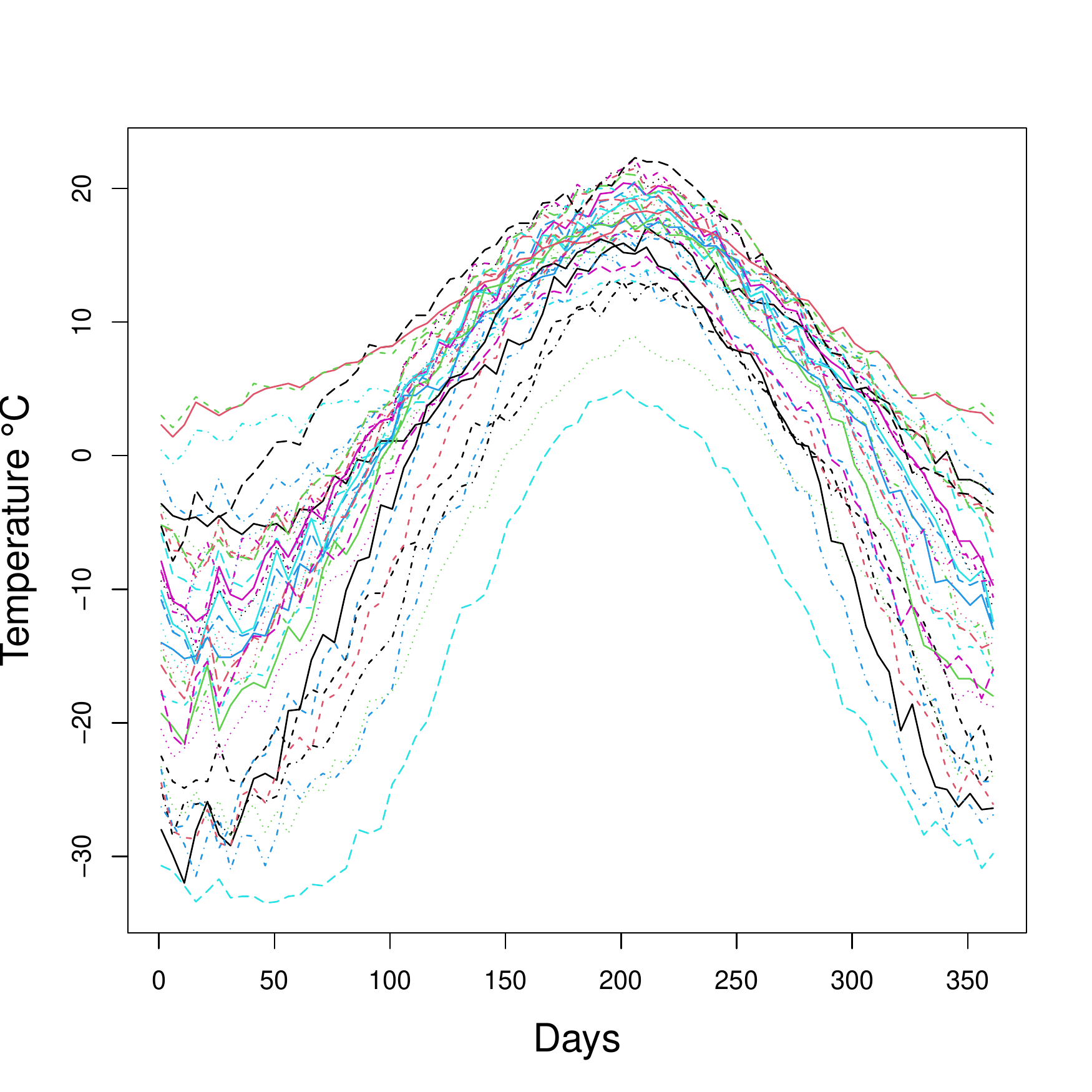}
 	\caption{ Daily mean temperature profiles at 35 cities in Canada over the year in the Canadian weather dataset.}
 \label{fig_candata}
 \end{figure}

The Canadian weather dataset contains the  daily mean temperature curves, measured in Celsius degree,   recorded at 35 cities in Canada. 
 The temperature   profiles are obtained by averaging over  the years 1960 through 1994.
 This is a benchmark dataset available in  the \textsf{R} package \textsf{fda} \citep{fda_package} that has been already studied by \cite{ramsay2005functional,centofanti2020smooth}.
	 Figure \ref{fig_candata} displays the interpolating profiles, where, for computational reasons, temperature curves are sampled each five days. 

 The ultimate goal of the cluster analysis applied to these curves is the geographical interpretation of the results.
 In particular,  all  methods analysed in Section \ref{sec_sim} are applied by setting $G=4$  in order to try to recover the grouping  of  4  climate
 zones, viz., Atlantic, Pacific, Continental, Arctic \citep{jacques2013funclust}.
The  first row of Table \ref{ta_can}  shows the $aRand$ index values of the resulting clusters calculated with respect to the 4-climate-zone grouping.
\begin{table}
	\caption{The values of the $aRand$ index for all the  clustering methods with respect to   climate
		zones grouping and the SaS-Funclust partition for the Canadian weather dataset \label{ta_can}}
	\begin{center}
		\resizebox{\textwidth}{!}{
			\begin{tabular}{ccccccccccc}
				\toprule
				&SaS-Funclust& curvclust& funHDDC&B-HC&B-KM&B-FMM&FPCA-KM&FPCA-HC&FPCA-FMM&DIS-KM\\
				\midrule
			 Climate zones grouping&0.37& 0.24& 0.21&  0.38  &0.21& 0.33 & 0.22& 0.30&0.17& 0.27\\
				SaS-Funclust&-&0.50& 0.35  &0.86 & 0.35 & 0.93 & 0.59& 0.72 & 0.43 &  0.40\\
				\bottomrule
			\end{tabular}
		}
	\end{center}
	
\end{table}
Although the SaS-Funclust and the B-HC  methods achieve the largest $aRand$   in this case,   $aRand$ values are in all cases inadequately low, which indicates  the clustering structure disagrees with such grouping.
That is, different method performance cannot properly evaluated by using the 4-climate-zone grouping.
The second row of Table \ref{ta_can} reports the $aRand$ index for all the competing methods  calculated with respect to the  SaS-Funclust method.
As expected, the proposed  clustering agrees  with filtering methods based on B-splines, while mostly  disagrees with the others.

\begin{figure}
	
	\centering
	\begin{subfigure}[b]{0.33\textwidth}
		
		\centering
	
	\includegraphics[width=\textwidth]{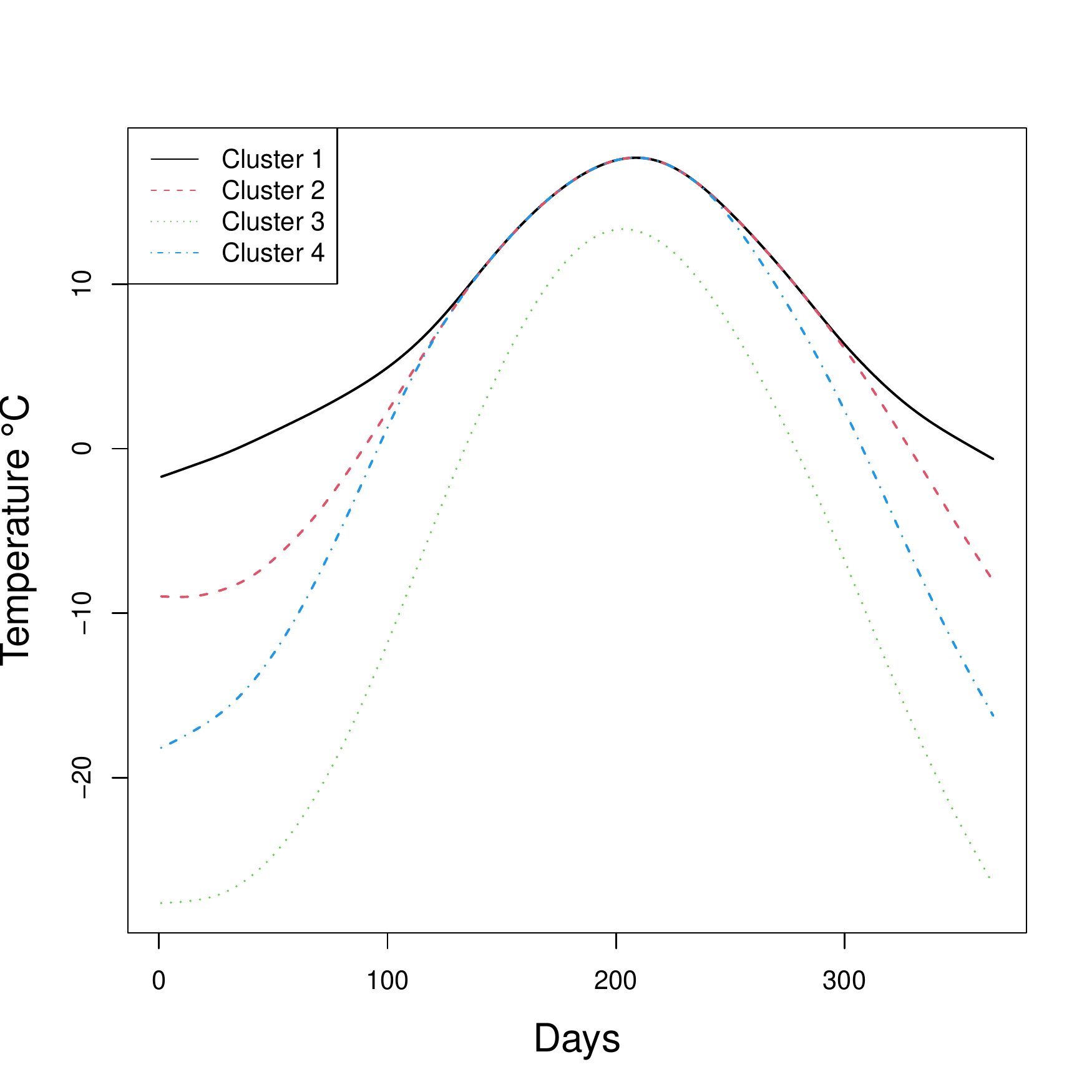}
	\vspace{-1.2cm}
	\caption{}
\end{subfigure}
	\centering
	\begin{subfigure}[b]{0.33\textwidth}
		
		\centering
	\includegraphics[width=\textwidth]{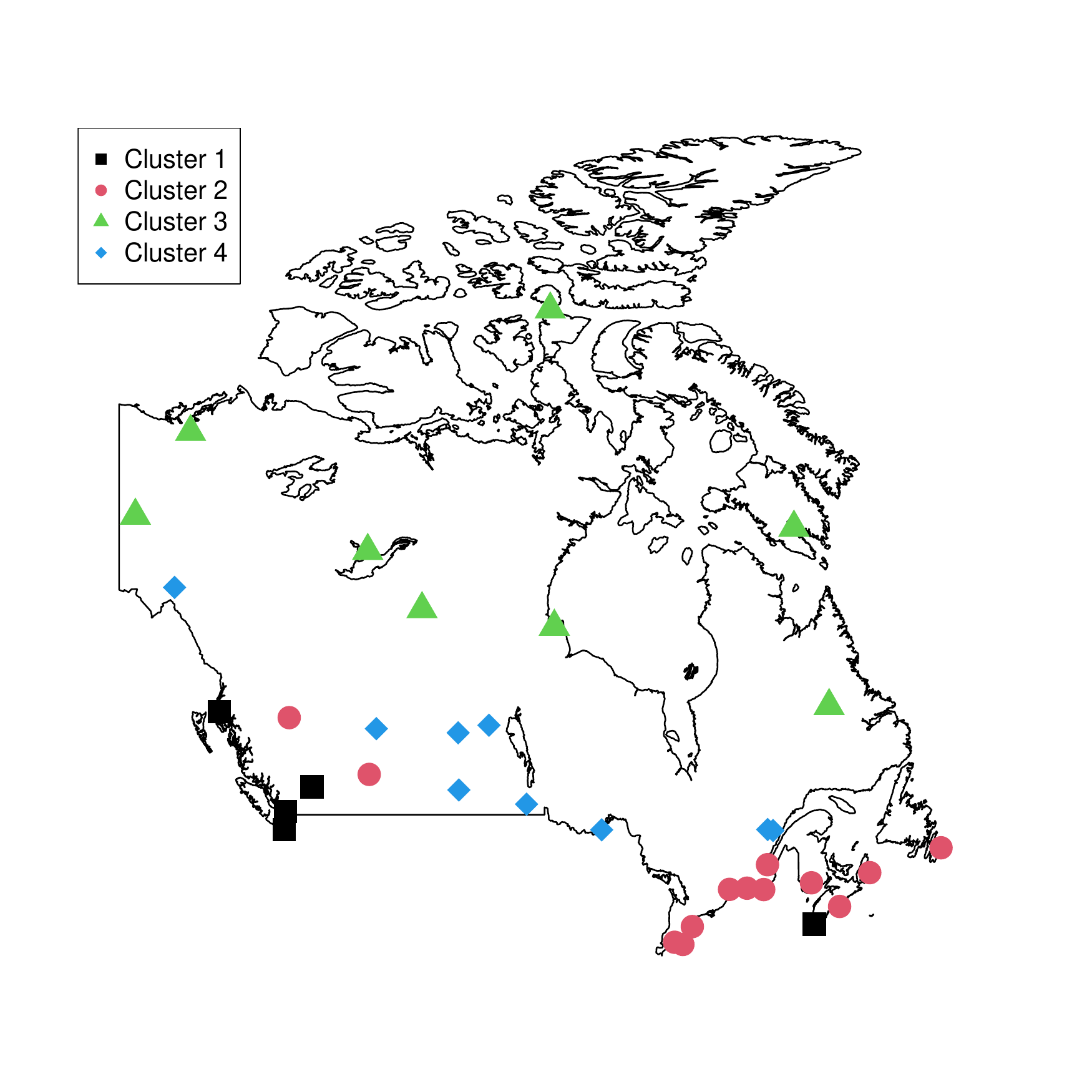}
	\vspace{-1.2cm}
	\caption{}
\end{subfigure}
	\caption{ (a) Estimated cluster curve mean and (b) geographical displacement of the curves pertaining to clusters obtained through  SaS-Funclust method.}
	\label{fig_can_mean}
	
\end{figure}

In terms of interpretability, Figure \ref{fig_can_mean} shows the estimated cluster  means and  geographical distribution of the curves in the clusters obtained by  the SaS-Funclust method.
From Figure \ref{fig_can_mean}(a), the estimated means for clusters 1, 2 and 4 are shown to fuse approximately from day 100 through 250. This is a strong evidence that the mean temperature in this period of the year is not significantly different among zones in cluster 1, 2 and 4. Hence, this portion of domain turns out to be noninformative for the separation of these clusters,
whereas the mean temperature is different for the rest of the year.
A different pattern is followed by the curves in cluster 3, which shows significantly smaller mean temperature all over the year.
The geographical displacement of the temperature profiles coloured by the clusters identified through the SaS-Funclust method is reported in Figure \ref{fig_can_mean}(b). Observations in cluster 1, 2 and 3 correspond to Pacific, Atlantic and southern continental stations and show similar mean temperature patterns only over the middle days of the year. Observations in cluster 3, which correspond to northern stations,  show lower mean temperature.
This nice and plausible interpretation  of this well-known real-data example is not possible by  means of any competing method.

\subsection{ICOSAF Project Data}
\begin{figure}
	
	\centering
	\includegraphics[width=0.33\textwidth]{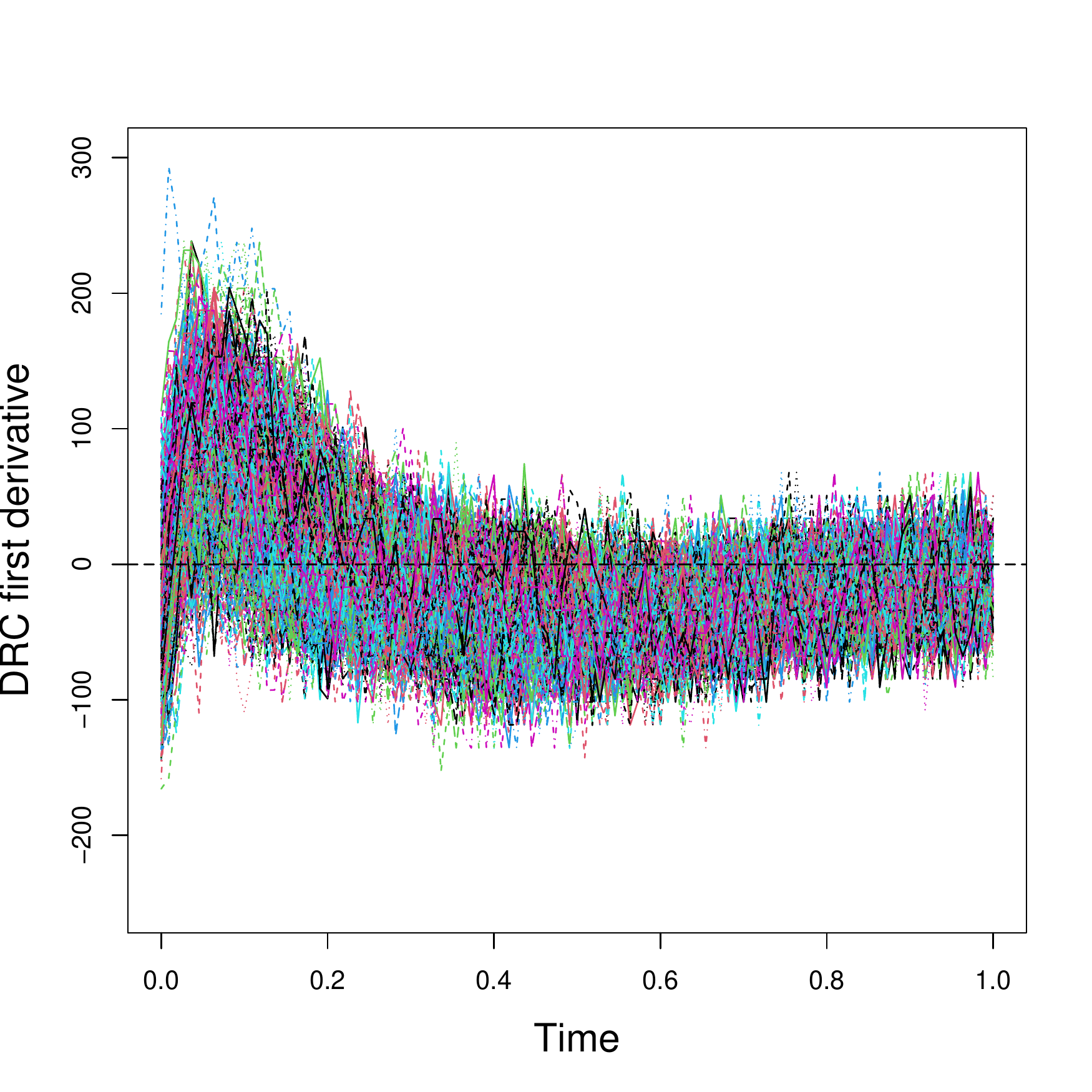}
	\caption{ First derivatives of the  538 DRCs in the ICOSAF project dataset.}
	\label{fig_ICOSAF}
\end{figure}

The ICOSAF project dataset contains  538 dynamic resistance curves (DRCs),   collected
during resistance spot welding lab tests at Centro Ricerche Fiat in 2019. The DRCs are collected over a regular grid of 238 points  equally spaced by 1 ms.
Further details on this dataset can be found in  \cite{capezza2020functional}  and the data are publicly available online at  \url{https://github.com/unina-sfere/funclustRSW/}.
In this example, we focus on the first derivative of the DRCs, estimated by means of the central differences method applied to the DRC values sampled each 2 ms.
Figure \ref{fig_ICOSAF} shows the  first derivative of the DRCs defined, without loss of generality, on the domain $ \left[0,1\right] $.
In this setting, the aim of the analysis is to cluster  DRCs to identify homogenous groups of spot welds that share common mechanical and metallurgical properties.
Differently from the previous datasets, no information are available about a reasonable partition of the DRCs.
	Therefore, based on the considerations provided by  \cite{capezza2020functional} and on cluster number selection methods described for  the  SaS-Funclust and competing methods in Section \ref{sec_modsel} and \ref{sec_sim}, respectively, we set $G=3$.
Table \ref{ta_ICOSAF} shows  $aRand$ values obtained  for all method pairs  with respect to the  SaS-Funclust partition. 
\begin{table}
	\caption{$aRand$ index calculated on the ICOSAF project dataset  for all competing method partitions with respect to    the SaS-Funclust one.  \label{ta_ICOSAF}}
	\begin{center}
		\resizebox{\textwidth}{!}{
			\begin{tabular}{ccccccccccc}
				\toprule
				& curvclust& funHDDC&B-HC&B-KM&B-FMM&FPCA-HC&FPCA-KM&FPCA-FMM&DIS-KM\\
				\midrule
				SaS-Funclust&	 0.00& 0.46& 0.41& 0.35& 0.46& 0.44& 0.27& 0.55&  0.56\\
				\bottomrule
			\end{tabular}
		}
	\end{center}
	
\end{table}
In this case, the  SaS-Funclust method provides partitions that are more similar to those obtained through the  FPCA-based methods than those obtained with the B-splines filtering approaches. 
However,  the clusters identified by the SaS-Funclust method  do not resemble those of the other methods.
It is worth noting that, for this dataset, even if results are not reported here, the	 partition obtained by curvclust differs dramatically  from the others and  does not provide  meaningful clusters.

\begin{figure}
	
	\centering
	\begin{subfigure}[b]{0.33\textwidth}
		
		\centering
		
		\includegraphics[width=\textwidth]{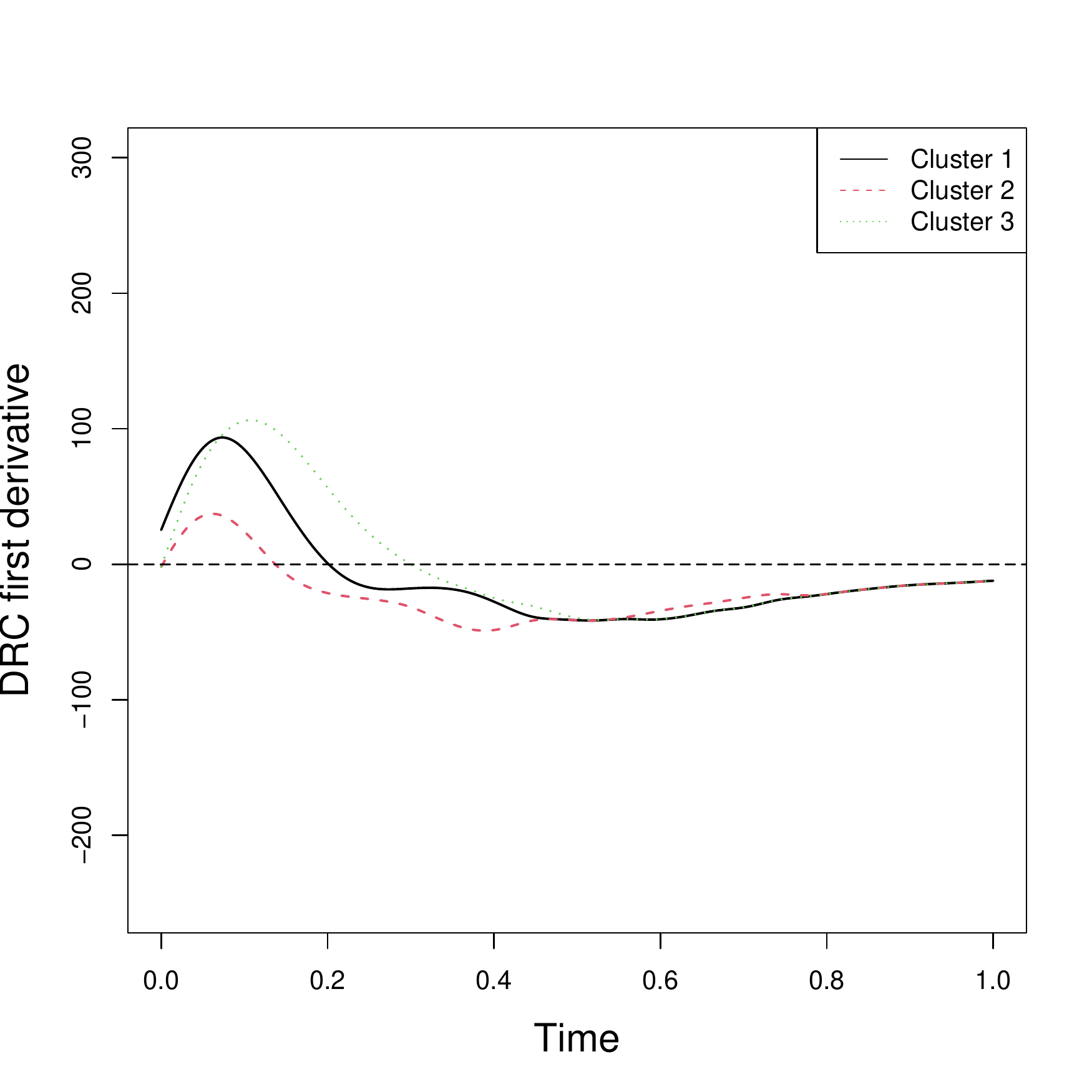}
		\vspace{-1.2cm}
		\caption{}
	\end{subfigure}
	\centering
	\begin{subfigure}[b]{0.33\textwidth}
		
		\centering
		\includegraphics[width=\textwidth]{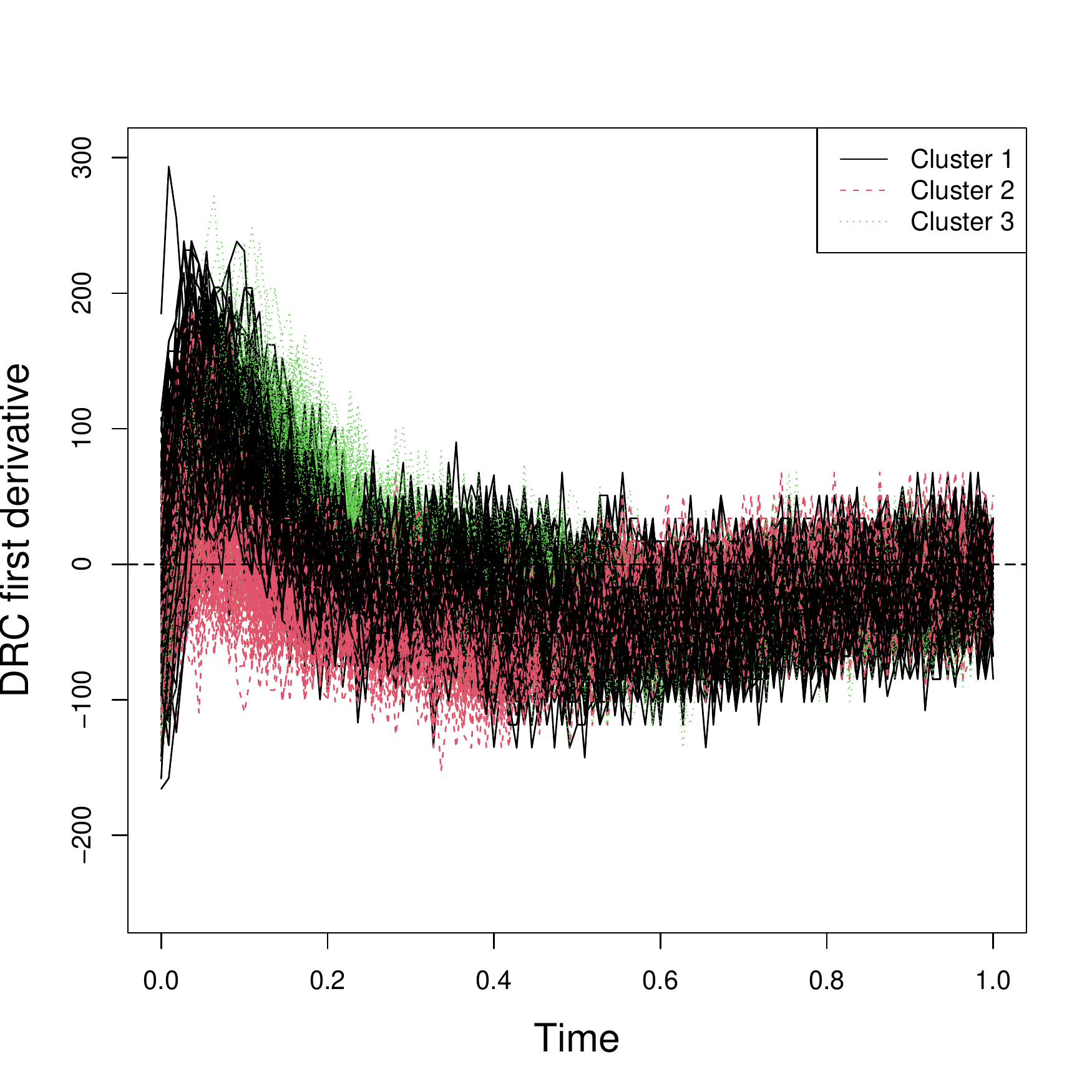}
		\vspace{-1.2cm}
		\caption{}
	\end{subfigure}
	\caption{ (a) Estimated cluster curve means and (b) curve clusters for the SaS-Funclust method  in the ICOSAF project dataset.}
	\label{fig_ICOSAF_mean}
	
\end{figure}

Also in this case, the SaS-Funclust method  allows for an insightful interpretation of the results.
The estimated cluster means and the corresponding clustered curves obtained through the SaS-Funclust method displayed in Figure \ref{fig_ICOSAF_mean} confirm the  ability of the proposed method  to  fuse cluster means, as it is clear over the second part of the domain. 
In particular, the mean of cluster 1 and 3 are fused from $ 0.5 $ to $ 1 $, which accounts for  the comparable decreasing rate of the DRCs over these clusters. Differently,  the mean of cluster 2 is fused with  other cluster means  between $ 0.8 $ and $ 1 $, only. This indicates that  between $ 0.5 $ and $ 0.8 $  DRCs of cluster 3 decrease  with a rate that is different from that of DRCs included in other clusters.
Differences between  cluster 2 and  clusters 1 and 3 are plainly visible  also in the first part of the domain, where DRCs of cluster 2 show   lower average velocity. Note also that  DRCs of cluster 2 reach their peaks (i.e., zeros of the first derivative)  earlier than those of clusters 3 and 1.

\section{Conclusions and discussions}
\label{sec:conc}
This article presented the  SaS-Funclust method, a new approach to the sparse clustering of functional data. 
Differently from methods that have already appeared in the literature before, it  was shown  to be capable of successfully detecting  where cluster pairs are separated.
In many applications, this  involves limited portions of domain, which are referred to as informative, and thus, the proposed method allows for a more accurate and interpretable cluster analysis.
The SaS-Funclust method can be considered as belonging to the  model-based clustering procedures with parameters of a general functional Gaussian mixture model   estimated by maximizing a penalized version of the log-likelihood function.
The key element  is the functional adaptive pairwise fusion penalty that,  by locally shrinking mean differences, allows pairs of cluster means to be exactly equal over  portions of domain where cluster pairs are not well separated, referred to as noninformative. In addition,  a smoothness penalty  is  introduced to further improve cluster interpretability.
The penalized log-likelihood function was maximized by means of a specifically designed expectation-conditional expectation algorithm, and model selection was addressed through a cross-validation technique.
An extensive Monte Carlo simulation study showed the favourable performance of the proposed method over several competing methods both in terms of clustering accuracy and interpretability. Lastly, real-data examples  further demonstrated the practical advantages  of the proposed method, which provided, thanks to its sparseness property, new insightful and interpretable solutions to cluster analysis.
In the  Berkeley growth study example, the SaS-Funclust method highlighted that growth velocity curves of boys and girls show different pubertal spurt, which happens later and last longer for male than female. Whereas, in the Canadian weather example, the mean temperatures over the  Pacific, Atlantic and southern
continental regions were found to be  equal over the middle days of the year and different otherwise. Moreover, the proposed method was applied to the ICOSAF project dataset, where, differently from the previous datsets, no information are available about a reasonable partition. The SaS-Funclust method also in this case identified homogenous groups of spot welds  that showed, only during the first part of the process,  differences in the rate of change of   dynamic resistance curves, which are likely to be  responsible of distinct  mechanical and metallurgical properties of the spot welds.

As closing remarks, we can envisage several important extensions  to  refine the proposed method.
Regarding the structure of the functional clustering model, the assumption of a  common  diagonal coefficient covariance matrix across  all clusters may  be too restrictive and may result in a poor fit.
Unfortunately, more flexible covariance structures  dramatically  increase the number of  parameters to be estimated,  already enlarged to achieve sparseness, in the  SaS-Funclust method. For this reason, regularization framework shall necessarily be addressed  to avoid overfitting, possibly either by constraining the covariance structure, as done in this article,  or by means of shrinkage estimators.
However, the choice of the best approach still remains not straightforward.
Furthermore, the covariance structure of the measurement errors  could be modified to include more complex relationships, and the model can be  extended also by including covariates \citep{james2003clustering}.

Another natural extension of the SaS-Funclust method that in worth considering is the integration of   a proper pairwise penalty applied to the covariance functions, useful in those settings where portions of domain  are informative for the clustering also in terms of covariance functions.
Unfortunately, the choice of such penalty and the resulting computational issues are  non-trivial and need for additional careful investigation. 

\subsection*{Acknowledgments}
The present work was developed within the activities of the project ARS01\_00861 ``Integrated collaborative systems for smart factory - ICOSAF'' coordinated by CRF (Centro Ricerche Fiat Scpa - \texttt{www.crf.it}) and financially supported by MIUR (Ministero dell’Istruzione, dell’Università e della Ricerca). 
We also acknowledge the CINECA award under the ISCRA initiative, for the availability of high performance computing resources and support.

\bibliographystyle{Chicago}

\small
\bibliography{References}

\end{document}


\def\spacingset#1{\renewcommand{\baselinestretch}%
{#1}\small\normalsize} \spacingset{1}


\if1\blind
{
  \title{\bf Supplementary materials for Sparse and Smooth Functional Data Clustering}
  \author[1]{Fabio Centofanti}
    \author[1]{Antonio Lepore \thanks{Corresponding author. e-mail: \texttt{antonio.lepore@unina.it}}}
     \author[1]{Biagio Palumbo}
  
  \affil[1]{Department of Industrial Engineering, University of Naples Federico II, Piazzale Tecchio 80, 80125, Naples, Italy}
  \date{}
  \maketitle
} \fi

\if0\blind
{
  \bigskip
  \bigskip
  \bigskip
  \begin{center}
    {\LARGE\bf Supplementary materials for Sparse and Smooth Functional Data Clustering}
\end{center}
  \medskip
} \fi

\bigskip

\setcounter{equation}{0}
\setcounter{figure}{0}
\setcounter{table}{0}
\setcounter{page}{1}
\makeatletter
\renewcommand{\theequation}{S\arabic{equation}}
\renewcommand{\thefigure}{S\arabic{figure}}
\renewcommand{\bibnumfmt}[1]{[S#1]}
\renewcommand{\citenumfont}[1]{S#1}

\section{Proof of Theorem 1}

We have that
\begin{align}
\label{eq_prof1}
	\sup_{t\in \mathcal{T}}|f\left(t\right)-\tilde{f}\left(t\right)|=&\sup_{t\in \mathcal{T}}|f\left(t\right)-\sum_{i=1}^{q}c_iI_{\left[\tau_i,\tau_{i+1}\right]}\left(t\right)|=\max_{i}\sup_{t\in\left[\tau_i,\tau_{i+1}\right]}|f\left(t\right)-c_i|.
\end{align}
Following \cite{de1978practical}, we have that for $ t \in\left[\tau_i,\tau_{i+1}\right]$ and $ i=1,\dots,q $,
\begin{equation}\label{eq_prof2}
|f\left(t\right)-c_i|\leq D\sup_{s\in\left[\tau_i,\tau_{i+1}\right]}|f\left(t\right)-f\left(s\right)|\leq D\sup_{s,z\in\left[\tau_i,\tau_{i+1}\right]}|f\left(z\right)-f\left(s\right)|\leq D\sup_{|s-t|\leq\delta_{\tau}}|f\left(t\right)-f\left(s\right)|,
\end{equation}
where $\delta_{\tau}=\max_{i}|\tau_{i+1}-\tau_{i}|$.
The term $ \sup_{|s-t|\leq\delta_{\tau}}|f\left(t\right)-f\left(s\right)| $ is the modulus of continuity of $f$ on $ \mathcal{T} $, and from \cite{schumaker2007spline}, the following inequality holds
\begin{equation}\label{eq_prof3}
\sup_{|s-t|\leq\delta_{\tau}}|f\left(t\right)-f\left(s\right)|\leq \delta_{\tau}\sup_{t\in\mathcal{T}}|f^{\left(1\right)}\left(t\right)|\leq\delta\sup_{t\in\mathcal{T}}|f^{\left(1\right)}\left(t\right)|,
\end{equation}
where the last inequality follows because $0\leq\delta_{\tau}\leq \delta $.
Therefore, upon using Equation \eqref{eq_prof1}, \eqref{eq_prof2}, and \eqref{eq_prof3}, it readily follows that
\begin{equation}\label{key}
\sup_{t\in \mathcal{T}}|f\left(t\right)-\tilde{f}\left(t\right)|\leq D \delta\sup_{t\in\mathcal{T}}|f^{\left(1\right)}\left(t\right)|,
\end{equation}
which proves the theorem.